\documentclass[11pt]{article}
\usepackage[margin=2cm]{geometry}
\usepackage{comment}
\usepackage{amsmath,amssymb,extarrows,mathtools,graphicx,subfigure,setspace}
\usepackage{cite}
\usepackage{slashed}
\usepackage{color}
\usepackage{amsmath,amsthm,amsfonts}

\usepackage{fullpage}
\makeatother

\numberwithin{equation}{section}

\def\bid{\hbox{1\hspace{-0.04in}I}} 

\newcommand{\be}{\begin{equation}}
\newcommand{\bea}{\begin{eqnarray}}
\newcommand{\eea}{\end{eqnarray}}
\newcommand{\ba}{\begin{array}}
\newcommand{\ea}{\end{array}}
\newcommand{\ee}{\end{equation}}

\def\g{\mathfrak{g}}
\def\ba{\begin{array}}
\def\ea{\end{array}}
\def\be{\begin{equation}}
\def\ee{\end{equation}}

\def\-1{^{-1}}

\input{xy}
\xyoption{all}
\xyoption{matrix}

\def\Aut{\mathrm{Aut}}

\def\g{\mathfrak{g}}

\theoremstyle{plain}

\date{}

\begin{document}
\onehalfspacing
\noindent
\begin{titlepage}
\hfill
\vspace*{20mm}
\vspace{-8mm}
\begin{center}
{\Large {\bf Lie superbialgebra structures on the Lie superalgebra $({\cal C}^3 + {\cal A})$ and  deformation of
related integrable Hamiltonian systems}\\
}

\vspace*{15mm} \vspace*{1mm} A. Eghbali \hspace{-1mm}{\footnote{a.eghbali@azaruniv.edu}}\hspace{1mm} and A. Rezaei-Aghdam \hspace{-2mm}{ \footnote{Corresponding author: rezaei-a@azaruniv.edu}}
\\

{\it  Department of Physics, Faculty of Basic Sciences,\\
 Azarbaijan Shahid Madani University, 53714-161, Tabriz, Iran \\ }

\vspace*{.4cm}

\vspace*{.4cm}

\end{center}
\begin{abstract}
Admissible structure constants related to the dual Lie superalgebras of particular Lie superalgebra $({\cal C}^3 + {\cal A})$ are found
by straightforward calculations from the matrix form of super Jacobi and mixed super Jacobi identities
which are obtained from adjoint representation. Then, by making use of the automorphism supergroup of the Lie superalgebra $({\cal C}^3 + {\cal A})$,
the Lie superbialgebra structures on the Lie superalgebra $({\cal C}^3 + {\cal A})$ are obtained and classified into inequivalent
31 families. We also determine all  corresponding coboundary
and bi-r-matrix Lie superbialgebras. The quantum deformations associated with some Lie superbialgebras $({\cal C}^3 + {\cal A})$ are obtained, together with
the corresponding deformed Casimir elements. As an application of these quantum deformations, we construct a deformed integrable Hamiltonian system from the representation of the Hopf superalgebra ${{U}_{_\lambda}}^{\hspace{-1mm}({\cal C}_{p=1}^{2,\epsilon} \oplus {\cal A}_{1,1})}\big(({\cal C}^3+{\cal A})\big)$.
\end{abstract}
\end{titlepage}

\section{\large Introduction}

From the mathematical standpoint, the study of Lie bialgebra structures provides
a primary classification of possible quantum deformations of a given Lie algebra.
 Many interesting examples of Lie bialgebras based
on complex semisimple Lie algebras have been given by Drinfeld
\cite{Drin}. In the case of simple Lie algebras, all Lie bialgebras of the coboundary type have been classified \cite{Belavin}
(see, also, \cite{Stolin}) and by using this
classification, all constant solutions of the classical Yang-Baxter
equation have been obtained. In the case of non-semisimple, only  some
of low-dimensional examples have been studied
\cite{{JR},{RHr},{Gomez}}.
A complete classification of Lie bialgebras with
reduction was given in \cite{Del}. However, a classification of
Lie bialgebras is out of reach, with similar reasons as for the classification of Lie
algebras.  On the other hand, from the physical standpoint, the theory of classical integrable systems
naturally relates to the geometry and representation theory of
Poisson-Lie groups and the corresponding Lie bialgebras and their
classical r-matrices (see, e.g., \cite{Kosmann}). In the same
way, Lie superbialgebras \cite{N.A}, as the underlying symmetry
algebras, play an important role in the integrable structure of
$AdS/CFT$ correspondence \cite{Bs}. Considering that there is a
universal quantization for Lie superbialgebras \cite{Geer}, one
can assign an important role to the classification of Lie
superbialgebras (especially low-dimensional Lie superbialgebras)
from both physical and mathematical point of view. Until now
there were distinguished and non-systematic ways for obtaining
low-dimensional Lie superbialgebras (see,  for instance,
\cite{{J.z},{J}}). We have recently presented a systematic way for obtaining and classifying  low-dimensional
Lie superbialgebras by using the adjoint representation of Lie
superalgebras \cite{ER1} and then have applied this method to classify the Lie superbialgebras $gl(1|1)$ \cite{geometry}.
In the present paper, we will try to perform the classification of the Lie superbialgebras $({\cal C}^3 + {\cal A})$. The reason of focusing on this classification
is that we have already begun a study on super Poisson-Lie symmetry in  WZW model based on the Lie supergroup $({ C}^3 + { A})$ \cite{ER8}.
In this way, we have shown that the dual model to the $({ C}^3 + { A})$ WZW model  is itself a WZW model on a
Lie supergroup whose Lie superalgebra is isomorphic to  $({\cal C}^3 + {\cal A})$.
However, to make that more completed, that is, to obtain a hierarchy of $({ C}^3 + { A})$ WZW models related to the
super Poisson-Lie T-duality, first we must obtain and classify all Lie superbialgebras $({\cal C}^3 + {\cal A})$. Moreover,
as explained in \cite{geometry},  it has been noticed that the Lie superalgebras $gl(1|1)$ and $({\cal C}^3 + {\cal A})$ are each the two-dimensional
Lie superbialgebras, i.e., they are isomorphic to four-dimensional Drinfeld superdoubles of the type $(2|2)$ \cite{ER4}.

On the other hand, it is well known that quantum groups as a new kind of symmetry related to the integrability of some
quantum models appeared in the context of quantum inverse scattering methods \cite{korepin}.
A direct and systematic method to construct $N$-particle completely
integrable Hamiltonian systems from representations of coalgebras with Casimir element
has been presented in Ref. \cite{Ballesteros}. Their construction shows that quantum deformations
can be interpreted as generating structures for  deformations of integrable Hamiltonian
systems with coalgebra symmetry. In this paper, we construct a deformed integrable Hamiltonian
system from a convenient representation of the quantum Lie superalgebra
$({\cal C}^3 + {\cal A})$ with the corresponding Casimir element.  This system is constructed
on a supersymplectic flat supermanifold of the superdimension-$(4|4)$ as the phase superspace.

In  section 2, we recall some properties of $Z_2$-graded vector
spaces and  basic definitions concerning Lie superbialgebras.  Section 3 is initiated by a representation of the
indecomposable  and decomposable Lie superalgebras of the type $(2 | 2)$ whose representatives are given by Tables 1 and 2, respectively.
Then, in order to obtain the Lie superbialgebra structures on $({\cal C}^3 + {\cal A})$, the automorphism supergroup
of $({\cal C}^3 + {\cal A})$ is derived at the end of this section. In section 4, we obtain the solutions
of the super Jacobi and mixed super Jacobi identities by making use of
the adjoint representations of the Lie superalgebras ${\g}$ and ${
\tilde \g}$, then we find 31 families of inequivalent Lie superbialgebra
structures on $({\cal C}^3 + {\cal A})$ whose representatives are classified
in Table 3. In section 5, we list coboundary
and bi-r-matrix Lie superbialgebras of $({\cal C}^3 + {\cal A})$ (triangular or quasi-triangular)
with their corresponding classical $r$-matrices in Table 4.
Making use of the Lyakhovsky and Mudrov formalism \cite{Lyakhovsky}, the Hopf superalgebras related to
some Lie superbialgebras $({\cal C}^3+{\cal A})$ of Table 3 are obtained as three Propositions in section 6.
As an application, we get at the end of this section a family of quantum integrable Hamiltonian
systems that can be constructed from a convenient representation of the Lie supercoalgebra  $({\cal C}^3+{\cal A})$ with the corresponding
Casimir element. The paper is closed by a final section that includes some remarks.

\vspace{4mm}
\section{\large Basic definitions and notation}

In order to make the paper somewhat self-contained, let us first recall some properties of $Z_2$-graded vector
spaces and Lie superbialgebras on an appropriate field.

If $V$ is a $Z_2$-graded vector space, then $V= V_{_B} \oplus V_{_F}$, and we refer to $ V_B$ and $V_F$ as the even and odd subspaces of $V$, respectively.
We define the gradation index $|~|: V\rightarrow \{0 , 1\}$ for the homogenous elements  of $V$ by \cite{footnote3}
\begin{eqnarray}\label{b.1}
|x|:= \left\{\begin{array}{ll}0&{\rm }
\;x \in V_B\\ 1&{\rm
} \;x \in V_F~.\end{array} \right.
\end{eqnarray}
The dual graded vector space $V^\ast =V_B^\ast \oplus V_F^\ast$ of $V$ inherits a natural $Z_2$-gradation.

{\it A Lie superalgebra} $\g$ is a $Z_2$-graded vector space, thus admitting the decomposition $\g =\g_{{_B}} \oplus \g_{_F}$, equipped with a bilinear
superbracket structure $[. , .]: \g \otimes \g \rightarrow \g$ satisfying the requirements of (graded) antisymmetry and super Jacobi identity.
In order to express them, it is useful  to introduce a basis in $\g$, $\{X_i\} \subset \g_{{_B}} \cup \g_{{_F}}$, and structure constants
\begin{eqnarray}\label{b.2}
[X_i , X_j]~=~f^k_{~ij} ~X_k.
\end{eqnarray}
Structure constants have to satisfy the following super Jacobi identity:
\begin{eqnarray}\label{b.3}
(-1)^{i(j+k)}{f^m}_{jl}{f^l}_{ki} + {f^m}_{il}{f^l}_{jk} +
(-1)^{k(i+j)}{f^m}_{kl}{f^l}_{ij}=0,
\end{eqnarray}
where
\vspace{-7mm}
\begin{eqnarray}\label{b.4}
{f^k}_{ij}=-(-1)^{ij}{f^k}_{ji},
\end{eqnarray}
and ${f^k}_{ij}=0$ whenever grade$(i)$ + grade$(j) \neq $ grade$(k)$.

We see that the superdimension of Lie superalgebra $\g$ is  $(m|n)$ if  $dim ~\g_{{_B}}=m$ and $dim ~\g_{{_F}}=n$.
It is also useful to define the supersymmetry bilinear form. A bilinear form $<.~,~.>$ on $\g$ is called supersymmetry if only if for any $x, y \in \g$
\begin{eqnarray}\label{b.5}
<x , y> = (-1)^{xy} <y , x>,
\end{eqnarray}
and it is called super ad-invariant if and only if
\begin{eqnarray}\label{b.6}
<[x , y] , z> = <x , [y , z]>,
\end{eqnarray}
for all $x, y, z \in \g$.

Let $\g ={\g}_{_B} \oplus {\g}_{_F}$ be a finite dimensional Lie superalgebra, and consider its dual $\g^\ast ={\g}_{_B}^\ast \oplus {\g}_{_F}^\ast$.
By definition, an element $x^\ast \in \g^\ast$ is a linear functional on $\g$, i.e., $x^\ast(y) = <x^\ast , y>$ for all $y \in \g$.
It is obvious that we can extend $<.~,~.>$ to $(\g \otimes \g)^\ast \otimes (\g \otimes \g)$ (in the present case, $(\g \otimes \g)^\ast =
\g^\ast  \otimes \g^\ast$) by setting
\begin{eqnarray}\label{b.7}
<x^\ast \otimes y^\ast , x \otimes y> = (-1)^{y^\ast~ x}  <x^\ast , x><y^\ast , y>,
\end{eqnarray}
for elements $x^\ast, y^\ast \in \g^\ast$  and $x, y \in \g$.

A {\it Lie superbialgebra} structure on a Lie
superalgebra $\g$ is a linear map $\delta :
\g \longrightarrow \g \otimes \g$, called  the super cocommutator, such that\\
(1)~~$\delta$ is a super one-cocycle on $\g$ with values in $\g \otimes \g$, i.e., for $x, y \in \g$,
\begin{equation}
\delta([x , y]) ~=~ [x \otimes 1 + 1 \otimes x ~,~ \delta(y)] - (-1)^{x y} [y \otimes 1 + 1 \otimes
y ~, ~\delta(x)].\label{b.8}
\end{equation}
(2) The dual map $[. , .]_{\ast}: {\g}^\ast \otimes {\g}^\ast \to {\g}^\ast$ defines a Lie superbracket on ${\g}^\ast$, i.e., a super skew-symmetric
bilinear map on $\g^\ast$ satisfying the super Jacobi identity. By definition, let us set
\begin{equation}
<[x^\ast , y^\ast]_{\ast}~ , ~ x>~ = ~<{\delta(x)} ~ ,~ x^\ast \otimes y^\ast>,\label{b.10}
\end{equation}
for $x^\ast, y^\ast \in \g^\ast$ and $x \in \g$.
The Lie superbialgebra defined in this way will be denoted by
$({\g} , {\g}^\ast)$ or $({\g},\delta)$ \cite {{N.A},{ER1}}.


Let $r$ be an element  of  ${\g} \otimes {\g}$. The super commutator given by
\begin{equation}\label{b.17}
\delta(x) = [x \otimes 1 + 1 \otimes x~ ,~ r],    ~~~~~~~~~~~x \in \g,
\end{equation}
defines a {\it coboundary} Lie superbialgebra if only if  $r$ fulfills the modified
graded classical Yang-Baxter equation (GCYBE)
\begin{equation}\label{b.18}
[x \otimes 1 \otimes 1 + 1 \otimes x \otimes 1 +  1 \otimes 1 \otimes x ~ ,~ [[r , r]]] = 0,   ~~~~~~~~~x \in \g,
\end{equation}
where the graded Schouten bracket is defined by
\begin{equation}
[[r , r]] := [r_{12}, r_{13}] + [r_{12} , r_{23}] + [r_{13} ,
r_{23}],\label{b.19}
\end{equation}
and, if $r=r^{ij}X_i \otimes X_j$, we have denoted  $r_{12}= r^{ij}X_i \otimes X_j \otimes 1$, $r_{13}=
r^{ij}X_i \otimes 1 \otimes X_j$ and $r_{23}= r^{ij}1 \otimes X_i
\otimes X_j$.
A solution of the GCYBE is often called a {\it
classical r-matrix} (in the following we call it an $r$-matrix). Using the fact that $r$ has even
Grassmann parity and Grassmann parity of $r^{ij}$ comes from
indices,  one can show that
$$
[r_{12},r_{13}]=(-1)^{i(k+l)+jl}\;r^{ij}r^{kl}\;[X_i,X_k] \otimes
X_j \otimes X_l,
$$
$$
~[r_{12},r_{23}]=(-1)^{(i+j)(k+l)}\;r^{ij}r^{kl}\;X_i \otimes
[X_j,X_k] \otimes X_l,
$$
$$
[r_{13},r_{23}]=(-1)^{i(k+l)+jl}\;r^{ij}r^{kl}\;X_i \otimes X_k
\otimes [X_j,X_l].
$$

Coboundary Lie superbialgebras can be of two different types: when the $r$-matrix is a super skew-symmetric solution  of the
 (GCYBE), so that $[[r , r]] = 0$, we shall say  the coboundary Lie superbialgebra is a {\it
triangular} one. In contrast, a super skew-symmetric solution $r$ of equation (\ref{b.18}) with non-vanishing
graded Schouten bracket
\begin{equation}
[[r , r]] = \omega,            \qquad \omega \in {\wedge}^3 {\g},\label{b.20}
\end{equation}
will give rise to a so-called {\it quasi-triangular} Lie superbialgebra
\cite{N.A}.
We note that if ${\g}$ is a Lie superbialgebra then
${\g}^\ast $ is also a Lie superbialgebra ,
but this is not always true for the coboundary property.

Suppose that $\g$ is a coboundary Lie superbialgebra with one-coboundary (\ref{b.17}) and
$\g^{\ast}$ be its dual  coboundary Lie superbialgebra with the one-coboundary
\begin{equation}\label{b.21}
\delta^\ast (x^\ast ) = [x^\ast  \otimes 1 + 1 \otimes x^\ast ~ ,~ r^\ast],
~~~~~~~~x^\ast  \in \g^\ast,~~~~~~~~~r^\ast \in  \g^{\ast} \otimes \g^{\ast},
\end{equation}
where $\delta^\ast: \g^{\ast} \rightarrow  \g^{\ast} \otimes \g^{\ast}$ is a  super one-cocycle on $\g^{\ast}$ defined by $r^{\ast}$. Then, we will call
the pair $(\g , {\g^{\ast}})$  a {\it bi- r-matrix  superbialgebra} \cite{RHr} if the graded
Lie brackets $[. ,  .]'$ on $\g$  defined by $\delta^\ast$
\begin{equation}\label{b.22}
<{\delta^\ast (x^\ast)} ~ ,~ x \otimes y> ~ = ~ <x^\ast ~ , ~ [x , y]'>,~~~~~~~~x, y  \in \g,~~~~~~~~~x^\ast \in  \g^{\ast},
\end{equation}
are equivalent to the original ones
\begin{equation}\label{b.22}
[x , y]' = A^{-1} [Ax  ,  Ay],~~~~~~~~x, y  \in \g,~~~~~~~~~A \in  Aut(\g),
\end{equation}
where $Aut(\g)$ stands for the automorphism  supergroup of the Lie superalgebra $\g$.


A {\it Manin supertriple}  is a triple of Lie
superalgebras $(\cal{D} , {\g} , {
\tilde{\g}})$ together with a nondegenerate ad-invariant
supersymmetric bilinear form (natural scalar product) $<.~ , ~.
>$ on $\cal{D}$, such that\hspace{2mm}

(1)~~${\g}$ and ${\tilde{\g}}$ are Lie
subsuperalgebras of $\cal{D}$,\hspace{2mm}

(2)~~$\cal{D} = {\g}\oplus{\tilde{\g}}$
as a supervector space,\hspace{2mm}

(3)~~${\g}$ and ${\tilde{\g}}$ are
isotropic with respect to the scalar product $< .~, ~. >$. By definition,  an isotropic
subspace is that the scalar product vanishes on it, i.e., for basis $\{X_i\} \in \g$ and $\{\tilde{X}^i\} \in \tilde \g$,
$$
<X_i , X_j> = <\tilde{X}^i , \tilde{X}^j> = 0,
$$
\begin{equation}
{\delta_i}^j=\;<X_i , \tilde{X}^j>\; = (-1)^{ij}<\tilde{X}^j,
X_i>=(-1)^{ij}{\delta^j}_i.\label{b.11}
\end{equation}
The Lie superbracket on $\tilde \g$ defines a Lie superbracket on $\g^{\ast}$. Also, to see it defines a Lie superbialgebra structure on $\g$,
we use the super Jacobi identity in $\cal{D}$ and the invariance of the scalar product. Thus,
there is a one-to-one
correspondence between Lie superbialgebra $({\bf \g},{\bf
\g}^\ast)$ and Manin supertriple $(\cal{D} , {\bf
\g} , { \tilde{\g}})$ with ${\bf
\tilde{\g}} \cong {\bf \g}^\ast$ \cite{Kosmann}.
Consider  the structure constants of Lie superalgebras ${\g}$ and
$\tilde{\g}$ as
\begin{equation}
[X_i , X_j] = {f^k}_{ij}~ X_k,\hspace{20mm} [\tilde{X}^i ,\tilde{
X}^j] ={{\tilde{f}}^{ij}}_{\; \; \: k}~ {\tilde{X}^k},\label{b.12}
\end{equation}
then, super ad-invariance of the bilinear form $< .~ , ~. >$ on $\cal{D}
= {\bf \g}\oplus{\bf \tilde{\g}}$ implies
\cite {ER1}
\begin{equation}
[X_i , \tilde{X}^j] =(-1)^j{\tilde{f}^{jk}}_{\; \; \; \:i}~ X_k
+(-1)^i {f^j}_{ki} ~\tilde{X}^k.\label{b.13}
\end{equation}
The Lie superbrackets (\ref{b.12}) and (\ref{b.13}) define a Lie superalgebra structure on the  vector space $\cal{D}$.
In this case, we say that the Lie superalgebra $\cal{D}$ is  the {\it Drinfeld superdouble } of $\g$ (or, equivalently, of ${ \tilde{\g}}$).
 In order to get an applicable result we use equations (\ref{b.7}), (\ref{b.10}), (\ref{b.11}) together with equation (\ref{b.12}) to obtain
\begin{equation}
\delta(X_i) = (-1)^{jk}{\tilde{f}^{jk}}_{\; \; \; \:i}~ X_j \otimes
X_k.\label{b.14}
\end{equation}
Utilizing this relation in the super one-cocycle
condition (\ref{b.8}), one can, respectively,  obtain  the super Jacobi identities  for the
dual Lie superalgebra and the  mixed super Jacobi
identities as follows:
\begin{equation}
(-1)^{i(j+k)}{\tilde{f}^{jl}}_{\; \; \; \; m}{\tilde{f}^{ki}}_{\; \; \; \; l}+
{\tilde{f}^{il}}_{\; \; \; \; m}{\tilde{f}^{jk}}_{\; \; \; \; l}+(-1)^{k(i+j)}{\tilde{f}^{kl}}_{\; \; \; \; m}{\tilde{f}^{ij}}_{\; \; \; \; l}=0,
\label{b.15}
\end{equation}
\begin{equation}
{f^m}_{jk}{\tilde{f}^{il}}_{\; \; \; \; m}=
{f^i}_{mk}{\tilde{f}^{ml}}_{\; \; \; \; \; j} +
{f^l}_{jm}{\tilde{f}^{im}}_{\; \; \; \; \; k}+ (-1)^{jl}
{f^i}_{jm}{\tilde{f}^{ml}}_{\; \; \; \; \; k}+ (-1)^{ik}
{f^l}_{mk}{\tilde{f}^{im}}_{\; \; \; \; \; j}.\label{b.16}
\end{equation}

\section{Lie superalgebras of the type $(2 | 2)$}

To present the notation  and for the self-consistency of the  paper, we use the list of
four-dimensional Lie superalgebras of the type $(2 | 2)$ of Ref. \cite{B}.
In that classification, Lie superalgebras are divided
into two types: trivial and nontrivial Lie superalgebras for which the fermion-fermion commutations are,
respectively, zero or non-zero. The results are presented in Table 1.
Because we use the DeWitt notation  and standard basis here,
the structure constants ${f^{^B}}_{_{FF}}$ must be purely imaginary \cite{D}.
Note that in \cite{B}, only the
indecomposable Lie superalgebras have classified. The decomposable Lie superalgebras of the type $(2 | 2)$ have been recently
obtained in \cite{geometry}, and here are presented in Table 2.
As can be seen from the Tables 1 and 2,
the Lie superalgebras have two bosonic generators $\{X_1, X_2\}$ and two fermionic ones
$\{X_3, X_4\}$. In labeling the trivial Lie superalgebras, the letters ${\cal A}, {\cal B}, {\cal C}$ and ${\cal D}$  denote the equivalence classes of Lie superalgebras of dimension $d$, where $d =
1, 2, 3$ and $4$, respectively, for ${\cal A}, {\cal B}, {\cal C}$ and ${\cal D}$. The superscript $i$ and real subscripts $p$ denote the respective number of
non-isomorphic Lie superalgebras and the Lie superalgebra parameter. For the nontrivial Lie superalgebras,
we add an integer superscript and a real subscript to parentheses around the symbol of the corresponding
trivial Lie superalgebra, where necessary.

Recently, we have classified all four-dimensional
Drinfeld superdoubles  of the type $(2 | 2)$ as a theorem in \cite{ER4}. We have shown that
there are just two classes of non-isomorphic Drinfeld superdoubles  of the type $(2 | 2)$ so that they are isomorphic to the
Lie superalgebras $gl(1|1)\big(\cong({\cal C}_{-1}^2 + {\cal A})\big)$ and $({\cal C}^3 + {\cal A})$ of Table 1.
So, four-dimensional Drinfeld superdoubles of the type $(2 | 2)$ have no new results
for Tables 1 and 2.
However, there are just 36 families of non-isomorphic  four-dimensional Lie superalgebras of the type $(2 | 2)$ which have presented in Tables 1 and 2.
We also note that some of the Lie superalgebras in the list contained at Table 1, such as $(2{\cal A}_{1,1}+2{\cal A})^2$,
${(2{\cal A}_{1,1}+2{\cal A})}^3_p$, and ${(2{\cal A}_{1,1}+2{\cal A})}^4_p$, can be considered to be relevant
for the $AdS_2/CFT_1$ correspondence. Because they, as the sub-superalgebras of the centrally-extended $psu(1|1)$ Lie superalgebra,
correspond to the algebra controlling the exact S-matrix theory of magnons transforming in a
centrally-extended $psu(1|1)$ superalgebra \cite{Torrielli1} (see, also, \cite{Torrielli2}).

\smallskip
\vspace{1mm}

\subsection{The  Lie superalgebra $({\cal C}^3+{\cal A})$ and its automorphism supergroup}

The  Lie superalgebra $({\cal C}^3+{\cal A})$ is spanned by the
set of generators $\{ X_1, X_2; X_3, X_4 \}$ with grading
$grade(X_1)=grade(X_2)=0$ and $grade(X_3)=grade(X_4)=1$, which in
the standard basis \cite{D} fulfill the following  (anti)commutation
relations \cite{B}
\begin{equation}
[X_1 , X_4] = X_3,\;~~~~~~\;\{X_4 , X_4 \}=iX_2,~~~~~~[X_2 ~, ~.] = 0,\;~~~~~[X_3 ~, ~.] = 0.\label{c.1}
\end{equation}
In  section 4, we will obtain  all the dual Lie superalgebras (the super cocommutators) related to the Lie superalgebra $({\cal C}^3+{\cal A})$. In this respect, we consider two Lie supercoalgebra structures $\delta$ and $\delta'$ equivalent if one can be obtained from the other by means of
a change of  basis which is an automorphism $A$ of the Lie superalgebra preserving the parity of the generators and the structure constants ${f^i}_{jk}$
($A: \g \rightarrow \g$).
Therefore it is crucial for our further considerations to obtain the automorphism supergroup of the particular
Lie superalgebra $({\cal C}^3+{\cal A})$.
\vspace{2mm}

 \hspace{-0.5cm}{\footnotesize Table 1.} {\small
{ Indecomposable Lie superalgebras of the type $(2 , 2).$}}\\
    \begin{tabular}{l l l  l p{15mm} }
    \hline\hline
{\scriptsize $ \g$ }& {\scriptsize Non-zero (anti)
commutation relations}&{\scriptsize Comments}  \smallskip\\
\hline
\smallskip

\vspace{-2mm}

{\scriptsize ${\cal D}^5~~$}& {\scriptsize $[X_1,X_3]=X_3, \;\;[X_1,X_4]=X_4,\;\;[X_2,X_4]=X_3$} \\

{\scriptsize ${\cal D}^6$}&{\scriptsize
$[X_1,X_3]=X_3, \;\;[X_1,X_4]=X_4,\;\;[X_2,X_3]=-X_4,\;\;[X_2,X_4]=X_3 $}\\

{\scriptsize ${\cal D}^1_{pq}$}&{\scriptsize
$[X_1,X_2]=X_2, \;\;[X_1,X_3]=pX_3,\;\;[X_1,X_4]=qX_4 $}& {\scriptsize $pq \neq 0,\;\; p\geq q$}\\

{\scriptsize ${\cal D}^8_{p}$}&{\scriptsize
$[X_1,X_2]=X_2, \;\;[X_1,X_3]=pX_3,\;\;[X_1,X_4]=X_3+pX_4 $}& {\scriptsize $p \neq 0$}\\

{\scriptsize ${\cal D}^9_{pq}$}&{\scriptsize
$[X_1,X_2]=X_2, \;\;[X_1,X_3]=pX_3-qX_4,\;\;[X_1,X_4]=qX_3+pX_4 $}& {\scriptsize $q > 0$}\\

{\scriptsize ${\cal D}^{10}_{p}$}&{\scriptsize
$[X_1,X_2]=X_2, \;\;[X_1,X_3]=(p+1)X_3,\;\;[X_1,X_4]=pX_4,\;\;[X_2,X_4]=X_3 $} \\

{\scriptsize ${({\cal D}^7_{\frac{1}{2}\;\frac{1}{2}})}^1$}& {\scriptsize
$[X_1,X_2]=X_2, \;\;[X_1,X_3]=\frac{1}{2}
X_3,\;\;[X_1,X_4]=\frac{1}{2}
X_4,\;\; \{X_3,X_3\}=iX_2,$} \\

& {\scriptsize
$\{X_4,X_4\}=iX_2$} \\

\vspace{-2mm}

{\scriptsize ${({\cal D}^7_{\frac{1}{2}\;\frac{1}{2}})}^2$}& {\scriptsize
$[X_1,X_2]=X_2, \;\;[X_1,X_3]=\frac{1}{2}
X_3,\;\;[X_1,X_4]=\frac{1}{2}
X_4,\;\; \{X_3,X_3\}=iX_2,$} \\

& {\scriptsize
$ \{X_4,X_4\}=-iX_2$} \\

{\scriptsize ${({\cal D}^7_{\frac{1}{2}\;\frac{1}{2}})}^3$}& {\scriptsize
$[X_1,X_2]=X_2, \;\;[X_1,X_3]=\frac{1}{2}
X_3,\;\;[X_1,X_4]=\frac{1}{2}
X_4,\;\; \{X_3,X_3\}=iX_2 $} \\

{\scriptsize $({\cal D}^7_{1-p\;p})$ }& {\scriptsize $[X_1,X_2]=X_2,
\;\;[X_1,X_3]=p X_3,\;\;[X_1,X_4]=(1-p)
X_4,\;\{X_3,X_4\}=iX_2 $}& {\scriptsize $p \leq \frac{1}{2}$} \\

{\scriptsize $({\cal D}^8_{\frac{1}{2}})$ }& {\scriptsize $[X_1,X_2]=X_2,
\;\;[X_1,X_3]=\frac{1}{2} X_3,\;\;[X_1,X_4]=X_3+\frac{1}{2}
X_4,\; \{X_4,X_4\}=iX_2 $} \\

\vspace{-1mm}

{\scriptsize $({\cal D}^9_{\frac{1}{2}\;p})$ }& {\scriptsize
$[X_1,X_2]=X_2, \;\;[X_1,X_3]=\frac{1}{2} X_3-p
X_4,\;\;[X_1,X_4]=p X_3+\frac{1}{2} X_4,$}&
 {\scriptsize $p > 0$}  \\

& {\scriptsize $\{X_3,X_3\}=iX_2,\;\; \{X_4,X_4\}=iX_2$}&
  \\

\vspace{-1mm}

{\scriptsize $({\cal D}^{10}_{0})^1$ }& {\scriptsize $[X_1,X_2]=X_2,
\;\;[X_1,X_3]= X_3, \;\;[X_2,X_4]= X_3, \;\; \{X_4,X_4\}=iX_1,$}\\

&{\scriptsize $\{X_3,X_4\}=-i\frac{1}{2}X_2$}\\

\vspace{-1mm}

{\scriptsize $({\cal D}^{10}_{0})^2$ }& {\scriptsize $[X_1,X_2]=X_2,
\;\;[X_1,X_3]= X_3, \;\;[X_2,X_4]= X_3, \;\; \{X_4,X_4\}=-iX_1,$}\\

&{\scriptsize $\{X_3,X_4\}=i\frac{1}{2}X_2$}\\

{\scriptsize $(2{\cal A}_{1,1}+2{\cal A})^2$}& {\scriptsize $\{X_3,X_3\}=iX_1,
\;\; \{X_4,X_4\}=iX_2, \;\; \{X_3,X_4\}=iX_1 $}&
 {\scriptsize Nilpotent}\\

{\scriptsize ${(2{\cal A}_{1,1}+2{\cal A})}^3_p$ }& {\scriptsize
$\{X_3,X_3\}=iX_1, \;\; \{X_4,X_4\}=iX_2, \;\;
\{X_3,X_4\}=ip(X_1+X_2)  $}
& {\scriptsize $p > 0\;$ Nilpotent}\\

{\scriptsize ${(2{\cal A}_{1,1}+2{\cal A})}^4_p$ }& {\scriptsize
$\{X_3,X_3\}=iX_1, \;\; \{X_4,X_4\}=iX_2, \;\;
\{X_3,X_4\}=ip(X_1-X_2)  $}
& {\scriptsize $p > 0\;$ Nilpotent}\\

{\scriptsize $({\cal C}^1_1+{\cal A})$ }& {\scriptsize $[X_1,X_2]=X_2,
\;\;[X_1,X_3]=X_3,
\;\;\{X_3,X_4\}=iX_2 $} \\

\vspace{-3mm}
{\scriptsize $({\cal C}^2_{-1}+{\cal A})$} & {\scriptsize $[X_1,X_3]=X_3,
\;\;[X_1,X_4]=-X_4,
\;\;\{X_3,X_4\}=iX_2 $}& {\scriptsize Jordan-Winger }\\

& & {\scriptsize quantization }\\

{\scriptsize $({\cal C}^3+{\cal A})$ }& {\scriptsize $[X_1,X_4]=X_3,
\;\;\{X_4,X_4\}=iX_2 $}&{\scriptsize Nilpotent}\\

{\scriptsize $({\cal C}^5_0+{\cal A})$ }& {\scriptsize $[X_1,X_3]=-X_4,
\;\;[X_1,X_4]=X_3,
\;\;\{X_3,X_3\}=iX_2,\;\;\{X_4,X_4\}=iX_2 $}\smallskip\\

\hline\hline
\end{tabular}\\

\vspace{2mm}
Since the Lie superalgebra is generated by the bosonic and fermionic generators, every automorphism is generated by
a linear transformation within the bosonic and fermionic sectors $\g_{{_B}}$ and $\g_{{_F}}$, respectively.
On the other hand, because of the preserving of grading the automorphisms of Lie superalgebra cannot mix fermionic with bosonic. Thus,
the action of the automorphism $A$ on $\g = \g_{_B} \oplus \g_{_F}$ is given by the block diagonal matrix  $A = diag(A_{_B} , A_{_F})$,
that is, in the standard basis \cite{D} $X'_i = \left(\begin{array}{c}
X'_{_B} \\ X'_{_F}  \end{array}
\right)$ and $X_i = \left(\begin{array}{c}
X_{_B} \\  X_{_F} \end{array} \right)$, we have the following transformation
\begin{equation}
X'_i = A(X_i) = (-1)^j  A_i^{~j}~  X_j,\label{c.2}
\end{equation}

\hspace{-0.5cm}{\footnotesize  Table 2. } {\small  Decomposable  Lie superalgebras of
the type $(2 , 2)$.}\\
    \begin{tabular}{l l l l  p{15mm} }
    \hline\hline
& {\scriptsize ${ \g}$ }& {\scriptsize Non-zero
commutation relations}&{\scriptsize Comments}  \smallskip\\
\hline
\smallskip
\vspace{-2mm}

{\scriptsize }&{\footnotesize ${\cal I}_{(2 , 2)}$}& {\scriptsize
All of the (anti)commutation relations are zero.} \\

&{\footnotesize ${\cal B} \oplus {\cal B}$}& {\footnotesize $[X_1,X_3]=X_3,
\;\;[X_2,X_4]=
X_4$} \\

&{\footnotesize ${\cal C}^1_{p} \oplus {\cal A}$}& {\footnotesize
$[X_1,X_2]=X_2, \;\;[X_1,X_3]=p X_3$} &{\footnotesize
$p \neq 0$}\\

&{\footnotesize ${\cal C}^2_{p} \oplus {\cal A}_{1,1}$}& {\footnotesize
$[X_1,X_3]=X_3, \;\;[X_1,X_4]=p X_4$} &{\scriptsize
$0< |p| \leq 1 $}\\

&{\footnotesize ${\cal C}^1_{p=0} \oplus {\cal A}$}& {\footnotesize
$[X_1,X_2]=X_2$}& \\

&{\footnotesize ${\cal B} \oplus {\cal A} \oplus {\cal A}_{1,1}$}& {\footnotesize
$[X_1,X_3]=X_3$}&{\scriptsize
$\equiv {\cal C}^2_{p=0} \oplus {\cal A}_{1,1}$} \\

&{\footnotesize ${\cal C}^3 \oplus {\cal A}_{1,1}$}& {\footnotesize
$[X_1,X_4]=X_3$}& {\scriptsize Nilpotent} \\

\vspace{0.5mm}

&{\footnotesize ${\cal C}^4 \oplus {\cal A}_{1,1}$}& {\footnotesize
$[X_1,X_3]=X_3,\;\;\; [X_1,X_4]=X_3+X_4$}& \\

\vspace{0.5mm}

&{\footnotesize ${\cal C}^5_p \oplus {\cal A}_{1,1}$}& {\scriptsize
$[X_1,X_3]=pX_3-X_4,\;[X_1,X_4]=X_3+pX_4$}&{\scriptsize
$p \geq 0 $} \\

\vspace{-1mm}

&{\footnotesize $(2{\cal A}_{1,1}+2{\cal A})^0$ }& {\footnotesize $\{X_3,X_3\}=iX_1  $}&
{\scriptsize $\equiv({\cal A}_{1,1}+{\cal A})\oplus {\cal A} \oplus {\cal A}_{1,1}$,}\\

&& &
{\scriptsize  Nilpotent}\\

\vspace{-1mm}

&{\footnotesize $(2{\cal A}_{1,1}+2{\cal A})^1$ }& {\footnotesize $\{X_3,X_3\}=iX_1,\; \;\; \{X_4,X_4\}=iX_2  $}&
{\scriptsize $\equiv({\cal A}_{1,1}+{\cal A})\oplus ({\cal A}_{1,1}+{\cal A})$,}\\
\vspace{-0.5mm}
&&&
{\scriptsize Nilpotent}\\

\vspace{0.5mm}

&{\scriptsize $({\cal B} \oplus ({\cal A}_{1,1}+{\cal A}))$ }& {\footnotesize $[X_1,X_3]=X_3,\; \;\; \{X_4,X_4\}=iX_2  $}\\

\vspace{0.5mm}

&{\scriptsize $(({\cal A}_{1,1}+2{\cal A})^1 \oplus {\cal A}_{1,1}) $ }& {\footnotesize
$\{X_3,X_3\}=iX_1,\; \;\; \{X_4,X_4\}=iX_1  $}&
 {\scriptsize Nilpotent}\\

\vspace{0.5mm}

&{\scriptsize $(({\cal A}_{1,1}+2{\cal A})^2 \oplus {\cal A}_{1,1}) $ }& {\footnotesize
$\{X_3,X_3\}=iX_1,\; \;\; \{X_4,X_4\}=-iX_1  $}&
 {\scriptsize Nilpotent}\\

\vspace{-1mm}

&{\footnotesize $({\cal C}^1_{\frac{1}{2}} + {\cal A})$ }& {\footnotesize
$[X_1,X_2]=X_2,\; \;[X_1,X_3]=\frac{1}{2}X_3,\; \;\{X_3,X_3\}=iX_2
$}&{\scriptsize $\equiv({\cal C}^1_{\frac{1}{2}}) \oplus {\cal A}$ }\smallskip \\

\hline\hline
\end{tabular}\\
\vspace{-1mm}
{\scriptsize  We recall that the Lie superalgebra ${\cal A}$ is  one-dimensional Abelian Lie superalgebra with one fermionic generator\\
while Lie superalgebra ${\cal A}_{1,1}$ is its bosonization.}\\

where $X'_i$ are the changed basis by the automorphism $A$. As mentioned above, the automorphism
preserves the structure constants, so the basis $X'_i$ must obey the same (anti)commutation relations as
$X_i$, i.e.,
\begin{equation}
[X'_i ,  X'_j] = {f^k}_{ij} ~X'_k.  \label{c.3}
\end{equation}
Inserting the transformation (\ref{c.2}) into (\ref{c.3}), we obtain the following matrix equation for the elements of
automorphism supergroup \cite{ER1}:
\begin{equation}
(-1)^{ij+mk} A~ {\cal Y}^k A^{st} = {\cal Y}^l
A_l^{\;\;k},   \label{c.4}
\end{equation}
where  $m$ denotes the column of
matrix $A^{st}$  in the left hand side, and the indices $i$ and $j$ correspond to the row and column of
matrix ${\cal Y}^k$, respectively. Here, $({\cal Y}^i)_{ \;jk} = -{f^i}_{jk}$ are the adjoint representations
of the Lie superalgebra $\g$. For $({\cal C}^3+{\cal A})$ one can use (\ref{c.1}) to get
{\small \begin{equation}
{\cal Y}^1 = 0,~~~~~ {\cal Y}^2 =\left(\begin{array}{cccc}
0 & 0  & 0 &  0\\
0 & 0  & 0 &  0\\
0 & 0  & 0 & 0\\
0 & 0  & 0 & -i
\end{array}
\right),~~~~~ {\cal Y}^3 =\left(\begin{array}{cccc}
0 & 0  & 0 &  -1\\
0 & 0  & 0 &  0\\
0 & 0  & 0 & 0\\
1 & 0  & 0 & 0
\end{array}
\right),~~~~~   {\cal Y}^4 = 0,\label{c.5}
\end{equation}}
then, using (\ref{c.4}) the automorphism $A$ can be easily obtained. The result is given by the following statement.

\vspace{2mm}

{\bf Proposition 1}: {\it The automorphism supergroup of the Lie superalgebra $({\cal C}^3+{\cal A})$
is expressed as matrices in basis $\{X_1, X_2; X_3, X_4\}$ as
{\small \begin{equation}\label{c.6}
 \Aut\big(({\cal C}^3+{\cal A})\big)=\left\{  A_{i}^{ ~j}=
\left(\begin{array}{cccc}
a & c   & 0 &  0\\
0 & b^2 & 0 &  0\\
0 & 0   & ab & 0\\
0 & 0   & d & b
\end{array}
\right);~~ a, b \neq 0
\right \}.
\end{equation}}}
\vspace{-3mm}
\section{The  Lie superbialgebra structures on $({\cal C}^3+{\cal A})$}
As mentioned in \cite{ER1}, because of tensorial form of super Jacobi and  mixed super Jacobi
identities (\ref{b.15}) and (\ref{b.16}), working with them is not so easy and we suggest
writing these equations as matrix forms using the following
adjoint representations for Lie superalgebras ${\g}$ and ${
\tilde \g}$
\begin{equation}
({\tilde{\cal X}}^i)^j_{\; \;k} = -{\tilde{f}^{ij}}_{\; \; \; k},
\hspace{10mm} ({\cal Y}^i)_{ \;jk} = -{f^i}_{jk}.\label{d.1}
\end{equation}
Then the matrix forms of identities (\ref{b.15}) and (\ref{b.16}) become, respectively, as follows:
\begin{equation}
({\tilde{\cal X}}^i)^j_{\; \;l}{\tilde{\cal X}}^l - {\tilde{\cal
X}}^j {\tilde{\cal X}}^i + (-1)^{ij}{\tilde{\cal X}}^i
{\tilde{\cal X}}^j = 0,\label{d.2}
\end{equation}
\begin{equation}
{({\tilde{\cal X}}^i)}^j_{\; \;l}\;{\cal Y}^l =-(-1)^{k}
({\tilde{\cal X}}^{st})^{j}\; {\cal Y}^i + {\cal Y}^j{\tilde{\cal
X}}^i - (-1)^{ij}{\cal Y}^i {\tilde{\cal X}}^j + (-1)^{k+ij}
({\tilde{\cal X}}^{st})^{i} \;{\cal Y}^j,\label{d.3}
\end{equation}
where index $k$ in the right hand side of equation (\ref{d.3}) represents the column of matrix ${\tilde{\cal
X}}^{st}$ and superscript $st$ stands for supertranspose.

To obtain the Lie superbialgebra structures on the Lie superalgebra  $({\cal C}^3+{\cal A})$, we must first solve
equations (\ref{d.2}) and (\ref{d.3}) in structure constants. The initial steps of the analysis  are made using a computer.
We recall that  ${\cal Y}^i$ in equation (\ref{d.3}) are
the adjoint representations of the Lie superalgebra $({\cal C}^3+{\cal A})$, which have presented in  (\ref{c.5}).
Thus by solving equations (\ref{d.2}) and (\ref{d.3}), we obtain the general form of the structure constants related to the  dual  Lie superalgebras to
$({\cal C}^3+{\cal A})$. The possibilities found by means of the computer are given by the following four cases:

\vspace{3mm}

 {\bf Theorem 1:} {\it The solutions of super Jacobi
and mixed super Jacobi identities \eqref{b.15} and \eqref{b.16} for the Lie superalgebra $({\cal C}^3+{\cal A})$ are given by the following four cases:}
\vspace{4mm}

\begin{tabular}{l l l l  p{15mm} }
\vspace{1.5mm}
{\it Case $(1):$ }& {  ${{\tilde{f}}^{12}}_{\; \; \: 1}= {{\tilde{f}}^{12}}_{\; \; \: 1},\;\;\;\;{{\tilde{f}}^{23}}_{\; \; \: 3}={{\tilde{f}}^{23}}_{\; \; \: 3},\;\;\;\;{{\tilde{f}}^{23}}_{\; \; \: 4}={{\tilde{f}}^{23}}_{\; \; \: 4},\;\;\;\;{{\tilde{f}}^{24}}_{\; \;
\: 4}={{\tilde{f}}^{12}}_{\; \; \: 1}+{{\tilde{f}}^{23}}_{\; \; \: 3}.$}   \smallskip\\

\vspace{1.75mm}

{\it Case $(2):$ }& {  ${{\tilde{f}}^{23}}_{\; \; \: 3}= {{\tilde{f}}^{23}}_{\; \; \: 3},\;\;\;\;{{\tilde{f}}^{24}}_{\; \; \:4}=-{{\tilde{f}}^{23}}_{\; \; \: 3},\;\;\;\;{{\tilde{f}}^{12}}_{\; \; \: 1}=-2{{\tilde{f}}^{23}}_{\; \; \: 3},\;\;\;\;{{\tilde{f}}^{23}}_{\; \;
\: 4}={{\tilde{f}}^{23}}_{\; \; \: 4},\;\;\;\;{{\tilde{f}}^{33}}_{\; \; \: 1}={{\tilde{f}}^{33}}_{\; \; \: 1}.$} \smallskip\\

\vspace{1mm}

{\it Case $(3):$ }& {  ${{\tilde{f}}^{33}}_{\; \; \: 1}= {{\tilde{f}}^{33}}_{\; \; \: 1},\;\;\;{{\tilde{f}}^{34}}_{\; \; \: 1}={{\tilde{f}}^{34}}_{\; \; \: 1},\;\;\;{{\tilde{f}}^{23}}_{\; \; \: 3}={{\tilde{f}}^{23}}_{\; \; \: 3},\;\;\;{{\tilde{f}}^{12}}_{\; \;
\: 1}=-{{\tilde{f}}^{23}}_{\; \; \: 3}+\frac{i}{2}{{\tilde{f}}^{34}}_{\; \; \: 1},\;\;\;{{\tilde{f}}^{24}}_{\; \; \: 4}=-\frac{i}{2}{{\tilde{f}}^{34}}_{\; \; \: 1},$}   \smallskip\\

\vspace{1.75mm}

& { ${{\tilde{f}}^{23}}_{\; \;
\: 4}=-\frac{{{\tilde{f}}^{33}}_{\; \; \: 1}(2 {{\tilde{f}}^{23}}_{\; \; \: 3} +i {{\tilde{f}}^{34}}_{\; \; \: 1})}{4{{\tilde{f}}^{34}}_{\; \; \: 1}}.$}   \smallskip\\

\vspace{1mm}

{\it Case $(4):$ }& {  ${{\tilde{f}}^{33}}_{\; \; \: 1}= {{\tilde{f}}^{33}}_{\; \; \: 1},\;\;\;\;{{\tilde{f}}^{34}}_{\; \; \: 1}={{\tilde{f}}^{34}}_{\; \; \: 1},\;\;\;\;{{\tilde{f}}^{33}}_{\; \; \: 2}={{\tilde{f}}^{33}}_{\; \; \: 2},\;\;\;\;{{\tilde{f}}^{12}}_{\; \;
\: 1}=\frac{i}{2}{{\tilde{f}}^{34}}_{\; \; \: 1},\;\;\;\;{{\tilde{f}}^{13}}_{\; \; \: 4}=\frac{i}{4}{{\tilde{f}}^{33}}_{\; \; \: 2},$}   \smallskip\\

\vspace{1.75mm}

& { ${{\tilde{f}}^{23}}_{\; \; \: 4}=-\frac{i}{4}{{\tilde{f}}^{33}}_{\; \; \: 1},\;\;\;\;{{\tilde{f}}^{24}}_{\; \; \: 4}=-\frac{i}{2}{{\tilde{f}}^{34}}_{\; \; \: 1}.$}   \smallskip\\

\end{tabular}\\

In the next step, we must find out that the above solutions are isomorphic with which one of the Lie superalgebras listed in Tables 1 and 2
(the detail of method has been explained in Ref. \cite{ER1}).
We have found that the solution of case (1) is isomorphic to the Lie superalgebras ${\cal C}^1_{p=\pm 1} \oplus {\cal A}$,
${\cal C}^2_{p=1} \oplus {\cal A}_{1,1}$,
${\cal C}^3 \oplus {\cal A}_{1,1}$, and ${\cal C}^4 \oplus {\cal A}_{1,1}$ of Table 2 and ${\cal D}^1_{p,q=p-1}$ of Table 1.
The solution of case (2) is isomorphic to the Lie superalgebras ${\cal D}^1_{p=\frac{1}{2},q=-\frac{1}{2}}$ and $({\cal C}^3 + {\cal A})$ of Table 1 and
 $(2{\cal A}_{1,1}+2{\cal A})^0$ of Table 2; moreover, this case is also
isomorphic to ${\cal C}^3 \oplus {\cal A}_{1,1}$ with the same conditions of case (1).
For case (3), we see that this solution
is isomorphic to the Lie superalgebras ${({\cal D}^7_{\frac{1}{2}\;\frac{1}{2}})}^2$, $({\cal D}^7_{1-p\;p})$,
$({\cal C}^1_1+{\cal A})$ and $({\cal C}^2_{-1}+{\cal A})$ of Table 1.
Finally we find that the solution of case (4) is isomorphic to the Lie superalgebras $({\cal D}^7_{1-p\;p})_{|_{p=0}}$,
$\big({\cal D}_{0}^{10}\big)^1$,  $\big({\cal D}_{0}^{10}\big)^2$ and $({\cal C}^3 +{\cal A})$ of Table 1. Also, case (4) is isomorphic to the Lie superalgebra
$({\cal C}^1_1+{\cal A})$ with the same conditions of  case (3).
Hitherto, we have found the general forms of the Lie superbialgebra structures on $({\cal C}^3+{\cal A})$. But now, one question raises: which
of these Lie superbialgebra structures are equivalent?
In order to answer this question, one has to use the automorphism supergroup of $({\cal C}^3+{\cal A})$ (\ref{c.6}) and the following proposition.

\vspace{2mm}

{\bf Proposition 2:} {\it If there exists an automorphism $A$ of ${\g}$ such that
\begin{equation}
\delta^\prime = (A\otimes A)\circ\delta\circ A^{-1},\label{d.4}
\end{equation}
then the super one-cocycles $\delta$ and $\delta^\prime$ of the
Lie superalgebra $\g$ are  equivalent. In this case, the
two Lie superbialgebras $({\g},\delta)$ and $({\g } ,\delta^\prime)$ are equivalent} \cite{Chari}.

From the action of both sides of relation (\ref{d.4}) on basis $\{X_i\}$ of $\g$  and then using relations
(\ref{c.2}) and (\ref{b.14}) we arrive at
\begin{equation}
(-1)^{jk+im} A^{-st}~ {\tilde {\cal Y} '}_{_i} A^{-1} = ({A^{-1}})_i^{~n} ~ {\tilde {\cal Y}}_{_n},\label{d.5}
\end{equation}
where  $m$ denotes the row of
matrix $A^{-st}$  in the left hand side, and the indices $j$ and $k$ correspond to the row and column of
matrix ${\tilde {\cal Y} '}_{_i}$, respectively.
Note that the bilinear form of the graded brackets (\ref{b.11}) is invariant with respect to
the transformations
$$
X'_i = (-1)^{l} A_{i}^{~~l}~ X_l,~~~~~~~~~~~{\tilde X}'^j = (A^{-st})^{j}_{~m}~ {\tilde X}^m.
$$
Inserting the second transformation into
$$
[{\tilde X}'^i  , {\tilde X}'^j] = {{\tilde f}}'^{ij}_{~~k} ~{\tilde X}'^k,
$$
one can recover relation (\ref{d.5}).
Finally we obtain 31 families of inequivalent Lie superbialgebra
structures on $({\cal C}^3 + {\cal A})$ whose representatives have classified
in Table 3.

\vspace{4mm}

\hspace{0.10cm}{\footnotesize  Table 3.} {\small  Dual Lie
superalgebras to the  Lie superalgebra $({\cal C}^3+{\cal A})$, $\epsilon=\pm 1$. }\\
\begin{tabular}{l l l   p{15mm} }
\hline\hline

{\footnotesize $ \tilde {\g}$ }& {\footnotesize Non-zero
(anti)commutation relations}&{\footnotesize Comments}  \smallskip\\
\hline
\smallskip

\vspace{0.5mm}

{\scriptsize ${\cal I}_{(2 , 2)}$}& {\scriptsize All of the (anti)commutation relations are zero.} \\

\vspace{1mm}

{\scriptsize ${\cal C}_{p=1}^{1,\epsilon} \oplus {\cal A}$}&{\scriptsize
$[\tilde X^1,\tilde X^2]= \epsilon \tilde X^1,\;\;\; [\tilde X^2,\tilde X^3]= -\epsilon \tilde X^3$}&\\

\vspace{1mm}

{\scriptsize ${\cal C}_{p=-1}^{1,\epsilon} \oplus {\cal A}$}&{\scriptsize
$[\tilde X^1,\tilde X^2]= \epsilon \tilde X^1,\;\;\; [\tilde X^2,\tilde X^4]= \epsilon \tilde X^4$}&\\

\vspace{1mm}

{\scriptsize ${\cal C}_{p=1}^{2,\epsilon} \oplus {\cal A}_{1,1}$}&{\scriptsize
$[\tilde X^2,\tilde X^3]= \epsilon \tilde X^3,\;\;\; [\tilde X^2,\tilde X^4]= \epsilon \tilde X^4$}&\\

\vspace{1mm}

{\scriptsize ${\cal C}^{3} \oplus {\cal A}_{1,1}.i$}&{\scriptsize
$[\tilde X^2,\tilde X^3]=  \tilde X^4$}&\\

\vspace{1mm}

{\scriptsize ${\cal C}^{4,\epsilon} \oplus {\cal A}_{1,1}$}&{\scriptsize
$[\tilde X^2,\tilde X^3]=  \tilde X^4,\;\;\; [\tilde X^2,\tilde X^3]= \epsilon \tilde X^3,\;\;\; [\tilde X^2,\tilde X^4]= \epsilon \tilde X^4$}&\\

\vspace{1mm}

{\scriptsize ${\cal D}_{p,p-1}^{1,\epsilon}$}&{\scriptsize
$[\tilde X^1,\tilde X^2]= -\epsilon \tilde X^1,\;\;\; [\tilde X^2,\tilde X^3]= \epsilon p \tilde X^3,\;\;\;
[\tilde X^2,\tilde X^4]= \epsilon(p-1) \tilde X^4$}& {\scriptsize $p \neq 0,1$}\\

\vspace{1mm}

{\scriptsize $\big({\cal D}_{\frac{1}{2},\frac{1}{2}}^{7,\epsilon}\big)^2$}&{\scriptsize
$[\tilde X^1,\tilde X^2]= -2\epsilon \tilde X^1,\;\;\; [\tilde X^2,\tilde X^3]= \epsilon  \tilde X^3,\;\;\;
[\tilde X^2,\tilde X^4]= \epsilon \tilde X^4,\;\;\;\{\tilde X^3,\tilde X^4\}= 2 i\epsilon \tilde X^1 $}& \\

\vspace{-1mm}

{\scriptsize $\big({\cal D}_{1-p,p}^{7,\epsilon}\big).i$}&{\scriptsize
$[\tilde X^1,\tilde X^2]= -\epsilon \tilde X^1,\;\;\; [\tilde X^2,\tilde X^3]= \epsilon p  \tilde X^3,\;\;\;
[\tilde X^2,\tilde X^4]= -\epsilon (p-1) \tilde X^4,$}&\\

&{\scriptsize $ \{\tilde X^3,\tilde X^4\}= -2 i\epsilon (p-1) \tilde X^1 $}& {\scriptsize $ p<\frac{1}{2},~p \neq 0$}\\

\vspace{-1mm}

{\scriptsize $\big({\cal D}_{1-p,p}^{7,\epsilon}\big).ii$}&{\scriptsize
$[\tilde X^1,\tilde X^2]= -\epsilon \tilde X^1,\;\;\; [\tilde X^2,\tilde X^3]= -\epsilon (p-1) \tilde X^3,\;\;\;
[\tilde X^2,\tilde X^4]= \epsilon p \tilde X^4,$}&\\

&{\scriptsize $ \{\tilde X^3,\tilde X^4\}= 2 i\epsilon p \tilde X^1 $}& {\scriptsize $ p<\frac{1}{2},~p \neq 0$}\\

\vspace{-1mm}

{\scriptsize $\big({\cal D}_{0}^{10,\epsilon}\big)^1$}&{\scriptsize
$[\tilde X^1,\tilde X^2]= -\epsilon \tilde X^1,\;\;\; [\tilde X^1,\tilde X^3]= -\frac{\epsilon}{4}  \tilde X^4,\;\;\;
[\tilde X^2,\tilde X^4]= \epsilon \tilde X^4, ~~~\{\tilde X^3,\tilde X^3\}=  i\epsilon \tilde X^2,$}&\\

&{\scriptsize $\{\tilde X^3,\tilde X^4\}= 2i\epsilon \tilde X^1$}&\\

\vspace{-1mm}

{\scriptsize $\big({\cal D}_{0}^{10,\epsilon}\big)^2$}&{\scriptsize
$[\tilde X^1,\tilde X^2]= -\epsilon \tilde X^1,\;\;\; [\tilde X^1,\tilde X^3]= \frac{\epsilon}{4}  \tilde X^4,\;\;\;
[\tilde X^2,\tilde X^4]= \epsilon \tilde X^4, ~~~\{\tilde X^3,\tilde X^3\}= - i\epsilon \tilde X^2,$}&\\

&{\scriptsize $\{\tilde X^3,\tilde X^4\}= 2i\epsilon \tilde X^1$}&\\

\vspace{1mm}

{\scriptsize $\big(2{\cal A}_{1,1} + 2 {\cal A} \big)^0.i$}&{\scriptsize
$\{\tilde X^3,\tilde X^3\}= i \tilde X^1$}&\\

\vspace{1mm}

{\scriptsize $\big({\cal C}_{1}^1 + {\cal A}\big)^\epsilon$}&{\scriptsize
$[\tilde X^1,\tilde X^2]= -\frac{\epsilon}{2} \tilde X^1,\;\;\;[\tilde X^2,\tilde X^4]= \frac{\epsilon}{2} \tilde X^4, ~~~
\{\tilde X^3,\tilde X^4\}= i\epsilon \tilde X^1$}&\\

\vspace{1mm}

{\scriptsize $\big({\cal C}_{-1}^2 + {\cal A}\big)^\epsilon$}&{\scriptsize
$[\tilde X^2,\tilde X^3]= {\epsilon} \tilde X^3,\;\;\;[\tilde X^2,\tilde X^4]=-{\epsilon} \tilde X^4, ~~~
\{\tilde X^3,\tilde X^4\}=-2 i\epsilon \tilde X^1$}&\\

\vspace{1mm}

{\scriptsize $\big({\cal C}^3 + {\cal A}\big)^\epsilon$}&{\scriptsize
$[\tilde X^1,\tilde X^3]= -\frac{\epsilon}{4}\tilde X^4,\;\;\;
\{\tilde X^3,\tilde X^3\}=i\epsilon \tilde X^2$}&\\

\vspace{1mm}

{\scriptsize $\big({\cal C}^3 + {\cal A}\big)_k^\epsilon$}&{\scriptsize
$[\tilde X^2,\tilde X^3]= {\epsilon} \tilde X^4,\;\;\;
\{\tilde X^3,\tilde X^3\}= ik \tilde X^1$}&{\scriptsize $k>0$}\smallskip\\

\hline\hline
\end{tabular}

\vspace{2mm}

\section{The coboundary Lie superbialgebras $({\cal C}^3+{\cal A})$}

The aim of this section is to find coboundary Lie superbialgebras of the Lie superalgebra $({\cal C}^3+{\cal A})$
with coboundary duals. As mentioned in section 2, since such structures can be specified (up to automorphism) by pairs of r-matrices,
so they are called bi-r-matrix Lie superbialgebras. For determining the coboundary Lie superbialgebras
of the Lie superalgebra $({\cal C}^3+{\cal A})$  we must find $r= r^{ij} X_i \otimes X_j \in ({\cal C}^3+{\cal A})\otimes({\cal C}^3+{\cal A})$.
To this end, one can use relations (\ref{b.2}) and (\ref{b.14}) to rewrite relation (\ref{b.17}) as
\begin{equation}\label{r.1}
{\tilde {\cal Y}}_i =  {{\cal X}_i}^{st}~r+ (-1)^l r {\cal X}_i,
\end{equation}
where the superscript $l$ corresponds to the row of the matrix ${\cal X}_i$. In this manner, we determine which of the
Lie superbialgebras ${\small ({\cal C}^3+{\cal A})}$ are coboundary and obtain the corresponding r-matrices.
On the other hand, by using the super one-cocycle (\ref{b.21} ) for some r-matrix ${\tilde r} \in {\tilde \g} \otimes {\tilde \g}$
and by considering ${\tilde \delta}({\tilde X}^i) = (-1)^{jk} f^i_{~jk} ~{\tilde X}^j \otimes {\tilde X}^k$ we get
\begin{equation}\label{r.3}
{{\cal Y}}^i =  {\tilde {\cal X}}^{i^{st}}~{\tilde r}+ (-1)^l {\tilde r} {\tilde {\cal X}}^i.
\end{equation}
Note that the super skew-symmetric
part of $r$ defined by $\hat{r}^{ij} =\frac{1}{2}(r^{ij} - (-1)^{ij} r^{ji})$ yields the same ${\tilde f}^{ij}_{\;~k}$, so we will assume
that $r \in \g \wedge \g$ with $r^{ij} = -(-1)^{ij} r^{ji}.$
In the same way, the mix part of $r$ as an element of $\g_{_B} \wedge \g_{_F}$ subspace of $\g \wedge \g$ cannot influence ${\tilde f}^{ij}_{\;~k}$ without violating the condition ${\tilde f}^{ij}_{\;~k} =0,$ if $i+j+k \neq 0$. Therefore, we get even $r$-matrix as $r \in \g_{_B} \wedge \g_{_B} \oplus \g_{_F} \wedge  \g_{_F}$, i.e., $r^{ij} =0$ if $i \neq j$.
To solve equations (\ref{r.1}) and (\ref{r.3}), we first obtain all adjoint representations of the Lie superbialgebras $({\cal C}^3+{\cal A})$ (listed in Table 3). Then  we find that the Lie superbialgebras $\big(({\cal C}^3+{\cal A}) , {\cal I}_{(2,2)} \big)$, $\big(({\cal C}^3+{\cal A}) , {\cal C}^3\oplus {\cal A}_{1,1}.i\big)$, $\big(({\cal C}^3+{\cal A}) , (2{\cal A}_{1,1} \oplus 2{\cal A}^{0}.i)\big)$ and
 $\big(({\cal C}^3+{\cal A}) , ({\cal C}^3+{\cal A})_k^\epsilon\big)$ are coboundary. Among these, only the Lie superbialgebras  $\big(({\cal C}^3+{\cal A}) , {\cal C}^3\oplus {\cal A}_{1,1}.i\big)$ and
 $\big(({\cal C}^3+{\cal A}) , ({\cal C}^3+{\cal A})_k^\epsilon\big)$  have coboundary duals. In the following, we give the solution of equations
 (\ref{r.1}) and (\ref{r.3}) for the coboundary  Lie superbialgebras.

{$\bullet$} For the Lie superbialgebra $\big(({\cal C}^3+{\cal A}) , {\cal I}_{(2,2)} \big)$, equation  (\ref{r.1})
can be solved and  the general solution
has the following form
\begin{eqnarray}
r = -a_1 \big(X_1 \otimes { X}_2 + X_2 \otimes { X}_1 - X_3 \otimes { X}_4 + X_4 \otimes { X}_3\big) + a_2X_2 \otimes { X}_2
+\frac{a_3}{2}X_3 \wedge { X}_3,\label{r.4}
\end{eqnarray}
where  $a_i$ $(i=1,2,3)$ are some the constant values. Note that, here and the following, the constants $a_i, b_i,...$
are  $c$-numbers \cite{D}.
For the above solution, the graded Schouten bracket is then read
\begin{eqnarray}
[[r , r]] = -\frac{{a_1}^2}{2}  X_2 \wedge { X}_3  \wedge { X}_3. \label{r.5}
\end{eqnarray}
Triangular solutions are obtained when ${a_1}$ and $a_2 $ are zero. As mentioned above, in this case, the  dual Lie superalgebra
is not coboundary.

{$\bullet$} The corresponding r-matrix to the coboundary Lie superbialgebra $\big(({\cal C}^3+{\cal A}) , {\cal C}^3\oplus {\cal A}_{1,1}.i\big)$ from
equation (\ref{r.1}) is obtained to be of the form
\begin{eqnarray}
r = X_1 \wedge { X}_2 -b_1 \big(X_1 \otimes { X}_2 + X_2 \otimes { X}_1 - X_3 \otimes { X}_4 + X_4 \otimes { X}_3\big) + b_2 X_2 \otimes { X}_2
+\frac{b_3}{2}X_3 \wedge { X}_3,\label{r.6}
\end{eqnarray}
for which the graded Schouten bracket is
\begin{eqnarray}
[[r , r]] = -\frac{{b_1}^2}{2} X_2 \wedge { X}_3  \wedge { X}_3. \label{r.7}
\end{eqnarray}
Then the condition $b_1= b_2 = 0$ gives rise to triangular solutions.
In this case we have a bi-r-matrix Lie superbialgebra, i.e., by solving equation (\ref{r.3}) we obtain an r-matrix $\tilde r \in {\tilde \g} \otimes {\tilde \g}$
$({\tilde \g}$ refers to ${\cal C}^3\oplus {\cal A}_{1,1}.i$) as
\begin{eqnarray}
\tilde r = -{\tilde X}^1 \wedge {\tilde X}^2 +c_1 {\tilde X}^1 \otimes {\tilde X}^1+ c_2 {\tilde X}^3 \otimes {\tilde X}^4 - (1+c_2){\tilde X}^4 \otimes {\tilde X}^3  +\frac{c_3}{2}  {\tilde X}^4 \wedge {\tilde X}^4,\label{r.8}
\end{eqnarray}
which induces on $\g = ({\cal C}^3+{\cal A})$ the original superbrackets (\ref{c.1}). For this solution, one can obtain
\begin{eqnarray}
[[\tilde r , \tilde r]] = \frac{1}{2} {\tilde X}^1 \wedge {\tilde X}^4 \wedge {\tilde X}^4, \label{r.9}
\end{eqnarray}
such that  quasi-triangular solutions are obtained when $c_1=0$ and $c_2=-\frac{1}{2}$.

{$\bullet$} The general solution of equation (\ref{r.1}) for the Lie superbialgebra $\big(({\cal C}^3+{\cal A}) , (2{\cal A}_{1,1} \oplus 2{\cal A}^{0}.i)\big)$
is given by
\begin{eqnarray}
r = -d_1 \big(X_1 \otimes { X}_2 - X_3 \otimes { X}_4\big)-(1+d_1)\big(X_2 \otimes { X}_1 + X_4 \otimes { X}_3\big) + d_2 X_2 \otimes { X}_2+\frac{d_3}{2} X_3 \wedge { X}_3,\label{r.10}
\end{eqnarray}
for which, we obtain
\begin{eqnarray}
[[r , r]] = -\frac{d_1(1+d_1)}{2} X_2 \wedge { X}_3  \wedge { X}_3. \label{r.11}
\end{eqnarray}
Therefore if $d_1 = -\frac{1}{2}$ and $d_2=0$,  we have quasi-triangular solutions.

{$\bullet$} For the Lie superbialgebra $\big(({\cal C}^3+{\cal A}) , ({\cal C}^3+{\cal A})_k^\epsilon\big)$, equations  (\ref{r.1}) and (\ref{r.3}) can be also solved and  the general solutions are, respectively, read
\begin{eqnarray}
r &=& (\epsilon-e_1) X_1 \otimes { X}_2-(\epsilon+e_1+k) X_2 \otimes { X}_1 +e_2 X_2 \otimes { X}_2 +\frac{e_3}{2} X_3 \wedge { X}_3~~~~~\nonumber\\
&&+e_1 X_3 \otimes { X}_4 - (k+e_1) X_4 \otimes { X}_3,\label{r.12}
\end{eqnarray}
\vspace{-7mm}
\begin{eqnarray}
\tilde r &=&f_1 {\tilde X}^1 \otimes {\tilde X}^1+{\epsilon}({-1+f_2 k}){\tilde X}^1 \otimes {\tilde X}^2  +(k+\frac{1}{\epsilon}+\frac{f_2 k}{\epsilon}){\tilde X}^2 \otimes {\tilde X}^1 + \frac{f_3}{2}  {\tilde X}^4 \wedge {\tilde X}^4 ~~~~~\nonumber\\
&&~+ f_2 {\tilde X}^3 \otimes {\tilde X}^4 - (\frac{1}{\epsilon}+f_2){\tilde X}^4 \otimes {\tilde X}^3.\label{r.13}
\end{eqnarray}
The corresponding graded Schouten brackets are then obtained to be
\begin{eqnarray}
[[r , r]] = - \frac{1}{2}(e_1^2 +e_1 k -\epsilon k) X_2 \wedge { X}_3  \wedge { X}_3 \label{r.14}
\end{eqnarray}
and
\begin{eqnarray}
[[\tilde r , \tilde r]] =\frac{1}{2}(\frac{1}{\epsilon} - k {f_{_2}}^2-\epsilon k f_2) {\tilde X}^1 \wedge {\tilde X}^4 \wedge {\tilde X}^4. \label{r.15}
\end{eqnarray}
The super skew-symmetric solutions of (\ref{r.12}) are given by $e_1 = -\frac{k}{2}$ and $e_2=0$. Thus, since $k$ is positive for $\epsilon =-1$
and $k=4$, we have triangular solutions, otherwise we are considering quasi-triangular ones. Similarly, by putting $f_1=0$ and
$f_2 = -\frac{1}{2 \epsilon}$ into (\ref{r.13}), we obtain the super skew-symmetric solutions. In the same way, triangular solutions are obtained only when
$\epsilon =-1$ and $k=4$.  We give the super skew-symmetry solutions of equations
(\ref{r.1}) and (\ref{r.3}) along with the corresponding graded Schouten brackets in Table 4.

\vspace{4mm}

\hspace{0.10cm}{\footnotesize  Table 4.} {\small  Super skew-symmetric r-matrix
solutions of the  $({\cal C}^3+{\cal A})$ Lie superbialgebras. }\\
\begin{tabular}{l l l l l  p{15mm} }
\hline\hline

{\footnotesize $ (\g , \tilde {\g})$ }&{\footnotesize $r, \tilde r$ } & {\footnotesize $[[r , r]], [[\tilde r , \tilde r]]$ }& \smallskip\\
\hline
\smallskip

\vspace{1mm}

{\scriptsize $\big(({\cal C}^3+{\cal A}) , {\cal I}_{(2,2)}\big)$}& {\scriptsize $ r = \frac{a_3}{2} X_3 \wedge { X}_3$} & {\scriptsize $[[r , r]] = 0$} &  & \\

\vspace{-1mm}

{\scriptsize $\big(({\cal C}^3+{\cal A}) , {\cal C}^3\oplus {\cal A}_{1,1}.i\big)$}& {\scriptsize $ r = X_1 \wedge { X}_2+ \frac{b_3}{2} X_3 \wedge { X}_3
$} & {\scriptsize $[[r , r]] = 0$} & & \\

\vspace{1mm}

& {\scriptsize $\tilde r = -{\tilde X}^1 \wedge {\tilde X}^2 -\frac{1}{2} {\tilde X}^3 \wedge {\tilde X}^4 +\frac{c_3}{2} {\tilde X}^4 \wedge {\tilde X}^4$} & {\scriptsize $[[\tilde r , \tilde r]] = \frac{1}{2} {\tilde X}^1 \wedge {\tilde X}^4 \wedge {\tilde X}^4 $} \\

\vspace{2mm}

{\scriptsize $\big(({\cal C}^3+{\cal A}) , (2{\cal A}_{1,1} \oplus 2{\cal A}^{0}.i)\big)$}& {\scriptsize  $r= \frac{1}{2}\big(X_1 \wedge { X}_2 -X_3 \wedge { X}_4 + d_3 X_3 \wedge { X}_3\big)
$} & {\scriptsize $[[ r ,  r]] = \frac{1}{8} X_2 \wedge { X}_3 \wedge { X}_3$} & & \\

\vspace{1mm}

\vspace{-1mm}

{\scriptsize $\big(({\cal C}^3+{\cal A}) , ({\cal C}^3+{\cal A})_k^\epsilon\big)$}& {\scriptsize $ r =(\epsilon+\frac{k}{2}) X_1 \wedge { X}_2+ \frac{e_3}{2} X_3 \wedge { X}_3-\frac{k}{2} X_3 \wedge { X}_4
$} & {\scriptsize $[[r , r]] = k(\epsilon+\frac{k}{4}) X_2 \wedge { X}_3  \wedge { X}_3$} & & \\

\vspace{1mm}

& {\scriptsize $\tilde r = -(\epsilon+\frac{k}{2}){\tilde X}^1 \wedge {\tilde X}^2-\frac{1}{2 \epsilon} {\tilde X}^3 \wedge {\tilde X}^4 + \frac{f_3}{2} {\tilde X}^4 \wedge {\tilde X}^4 $} & {\scriptsize $[[\tilde r , \tilde r]] =\frac{1}{2} (\epsilon+\frac{k}{4}){\tilde X}^1 \wedge {\tilde X}^4 \wedge {\tilde X}^4$} \\\hline\hline

\end{tabular}
\vspace{1mm}
\section{The quantum deformation of $({\cal C}^3+{\cal A})$ and deformation of
related integrable Hamiltonian systems}

In this section, we quantize the  Lie superalgebra  $({\cal C}^3+{\cal A})$ by making use of
the Lyakhovsky and Mudrov formalism \cite{Lyakhovsky}. For this
purpose and self-containing of the paper, we first review the main
result of this formalism as a following proposition. Then,  we use
this method in order to build up the Hopf superalgebras related to
some Lie superbialgebras $({\cal C}^3+{\cal A})$ of Table 3. Nevertheless,
as an application, we get at the end of this section a family of quantum integrable Hamiltonian
systems that can be constructed from a convenient representation of the quantum Lie superalgebra  $({\cal C}^3+{\cal A})$ with the corresponding
Casimir element.

\subsection{The Hopf superalgebras corresponding to some Lie superbialgebras $({\cal C}^3+{\cal A})$}
{\bf Proposition 3:} \cite{Lyakhovsky} {\it
Let  $\bid$, $H_i~(i=1, \cdots, n)$ and $\mathbb{X}_m~(l=1, \cdots, m)$ be a basis of an associative unital algebra
${\cal A}$ over the field $\mathbb{C}$, and  $\mu_i$, $\nu_j ~(i, j = 1, \cdots ,n)$ be a
set of $m \times m$ complex matrices such that they are commute
with together.
In addition, let ${\overrightarrow {\mathbb{X}}}$ be a column(row) vector with component $\mathbb{X}_l
(l= 1, \cdots ,m)$. The {\it coproduct}
\begin{eqnarray}
\triangle({\overrightarrow {\mathbb{X}}}) &=& \exp(\sum^{n}_{i=1}
{\mu_i}{H_i}) \dot{\otimes} {\overrightarrow {\mathbb{X}}} + \sigma
\Big(\exp(\sum^{n}_{i=1} {\nu_i}{H_i}) \dot{\otimes}
{\overrightarrow {\mathbb{X}}}\Big)\nonumber\\
\bigtriangleup(H_i)  &=&  \bid \otimes H_i + H_i \otimes \bid,\nonumber\\
\triangle(\bid) & = & \bid \otimes \bid,\label{Quantiz.1}
\end{eqnarray}
{\it the counit}
\begin{equation}
 \epsilon(H_i)\;=\; \epsilon
(\mathbb{X}_l)\;=\;0,  \qquad  \epsilon(\bid)\;=\;\bid, \label{Quantiz.2}
\end{equation}
{\it and the antipode}
\vspace{-1mm}
\begin{eqnarray}
\gamma(H_i) & =& -H_i, \nonumber\\
\gamma({\overrightarrow {\mathbb{X}}}) & =& - {\overrightarrow {\mathbb{X}}} \exp(\sum^{n}_{i=1}
{-\mu_i}{H_i}) \exp(\sum^{n}_{i=1}
{-\nu_i}{H_i}), \nonumber\\
\gamma(\bid) & =& \bid,\label{Quantiz.3}
\end{eqnarray}
endow $({{\cal A}}, \Delta, \epsilon, {\cal \gamma})$  with a
Hopf algebra structure if the generators $H_i$ commute
with together.}

Let $\bf P$ be an $m \times m$ matrix with entries $p_{_{kl}} \in {\cal A}$, then the $k$th component
of ${\bf P} \dot{\otimes} {\overrightarrow {\mathbb{X}}}$ is defined as
$({\bf P} \dot{\otimes} {\overrightarrow {\mathbb{X}}})_k = \sum^{m}_{l=1} p_{_{kl}} \otimes \mathbb{X}_l$.
Also, notice that here $\sigma$ is the flip operator $\sigma(\mathbb{X}_l \otimes \mathbb{X}_m) =  \mathbb{X}_m \otimes \mathbb{X}_l$.
The deformation parameters in the resulting coalgebra are the
entries of the matrices $\mu_i$ and $\nu_j$. If we can find a
compatible multiplication with the coproduct \eqref{Quantiz.1} we will have finally obtained a quantum
algebra. In fact, the role of the matrices $\mu_i$ and $\nu_j$ is to reflect the
Lie bialgebra underlying a given quantum deformation. This can be clearly appreciated by
taking the first order  of \eqref{Quantiz.1}
\begin{equation}
\bigtriangleup_{(1)}({\overrightarrow {\mathbb{X}}})\;=\;(\sum^{n}_{i=1}
{\mu_i}{H_i}) \dot{\otimes} {\overrightarrow {\mathbb{X}}} + \sigma
(\sum^{n}_{i=1} {\nu_i}{H_i} \dot{\otimes} {\overrightarrow
{\mathbb{X}}}).\label{Quantiz.4}
\end{equation}
On the other hand, since the cocommutator $\delta$ corresponds to the co-antisymmetric part of \eqref{Quantiz.4}, it can be written as
\begin{equation}
\delta ({\overrightarrow
{\mathbb{X}}})\;=\;\bigtriangleup_{(1)}({\overrightarrow {\mathbb{X}}})-\sigma \circ
\bigtriangleup_{(1)}({\overrightarrow {\mathbb{X}}}).\label{Quantiz.5}
\end{equation}
In this formalism, elements $H_i$ are called  primitive
generators. These elements must be chosen  such that $\delta
({\overrightarrow {\mathbb{X}}}_i)$ does not contain terms of the form $H_i
\wedge H_j$. We note that the same cocommutator (\ref{Quantiz.5})
can be obtained from different choices of the matrices $\mu_{i}$
and $\nu_{j}$, i.e., the different sets of matrices  lead to right
quantization of $U({\g})$.
When the algebra ${\cal A}$ is a Lie algebra ${\g}$, then
$H_i$ generate an Abelian Lie subalgebra, and with this
condition the deformed commutation relations in
$U_{_\lambda}({\g})$ with $\lambda$ being the deformation parameter are given by \cite{Lyakhovsky}
\begin{equation}
[\mathbb{X}_l , \mathbb{X}_p]\;=\;[\mathbb{X}_l , \mathbb{X}_p]_{\circ} + \phi_{lp}(\mu_i, \nu_j,
H_k),\label{Quantiz.6}
\end{equation}
where $[\mathbb{X}_l , \mathbb{X}_p]_{\circ}$ is the classical commutation relation
and the deforming functions $\phi_{lp}$  are the power series of
$H_k$'s.  Note that after determining of $\phi_{lk}$, the Jacobi
identity for (\ref{Quantiz.6})  must be checked.

The above formalism was presented for the Lie algebras and one can
use this formalism for Lie superalgebras by keeping that the
graded tensor product law must be taken into account \cite{kulish}
\begin{equation}
(F \otimes G)_{ij;kl}\;=\;(-1)^{j(i+k)}\;
F_{ik}G_{jl}.\label{Quantiz.7}
\end{equation}
If $U(\g)$ be a quantum superalgebra, the extension of the coproduct $\Delta: U(\g)\rightarrow U(\g) \otimes U(\g)$ to products of generators
should be substituted by $(a \otimes b)(c \otimes d)=(-1)^{bc} ac \otimes bd$ for all $a,...,d$ in $U(\g)$;
moreover, the flip operator $\sigma$ should be written as $\sigma(a \otimes b) = (-1)^{ab} b \otimes a$.

This quantization procedure can be applied to the  Lie
superbialgebras of Table 3 to quantize  the  Lie
superalgebra $({\cal C}^3+{\cal A})$.
We shall write  the supercoproduct and the deformed (anti)commutation
rule, as the supercounit is always trivial and the superantipode can be easily deduced by means
of the Hopf superalgebra axioms. Deformed Casimir operator, which is essential for the
construction of integrable systems, is also explicitly given.

Let us first denote the Lie superalgebra  $({\cal C}^3+{\cal A})$ with the generators $Z, H, Q_+$ and $Q_-$ instead of $X_1, X_2, X_3$ and $X_4$, respectively.
Then, in the non-standard basis, the  (anti)commutation
relations \eqref{c.1} are written as
\begin{equation}
[Z , Q_-]\;=\;Q_+,~~~~~~~\{Q_- , Q_-\}\;=\;H,~~~~~~~
[{H} ~, ~.]\;=\;0,~~~~~~~[{Q}_+~ , ~.]\;=\;0.
 \label{Quantiz.8}
\end{equation}
The relevant invariant supersymmetric bilinear form for $({\cal C}^3+{\cal A})$ is
\begin{equation}
<Z , H> ~=~ <H , Z> ~=~ <Q_- , Q_+> ~=~ - <Q_+ , Q_->~=~ 1.
 \label{Quantiz.9}
\end{equation}
For the above choice of  metric, the quadratic Casimir is found to be of the form
\begin{equation}
{\cal C}^{^{(2)}} ~=~ 2 (Z H - Q_+  Q_-).
 \label{Quantiz.10}
\end{equation}
Now we proceed to obtain the Hopf superalgebra corresponding to the Lie superbialgebra
$\big(({\cal C}^3+{\cal A}) , {\cal C}_{p=1}^{2,\epsilon} \oplus {\cal A}_{1,1}\big)$.
Commutation relations of the dual Lie superalgebra ${\cal C}_{p=1}^{2,\epsilon} \oplus {\cal A}_{1,1}$ have displayed in
Table 3. Using relation \eqref{b.14}, one can write the super cocommutators  as
\begin{eqnarray}
\delta(Z) &=& 0,~~~~~~~~~~~~~~~~~~~~ ~\delta(H) ~=~ 0, \nonumber\\
\delta(Q_+) &=& \epsilon H \wedge Q_+,~~~~~~~~~~
\delta(Q_-) ~=~ \epsilon H \wedge Q_-.\label{Quantiz.11}
\end{eqnarray}
We see that there does not exist the term of the type $Z \wedge H$ within the super cocommutators \eqref{Quantiz.11} and $[Z , H] =0$. So,
we denote the primitive generators $H_i$ by
$H_1 = Z$ and $H_2 =H$. Hence, by considering $\epsilon = \lambda$ the super cocommutators for the non-primitive generators $Q_+$ and $Q_-$ can be written as
\begin{eqnarray}
\delta \left(\begin{array}{cc}
Q_+    \\
Q_-
\end{array}
\right)~=~ \left(\begin{array}{cc}
\lambda H & 0   \\
0 & \lambda H
\end{array}
\right) \dot{\wedge} \left(\begin{array}{cc}
Q_+    \\
Q_-
\end{array}
\right). \label{Quantiz.12}
\end{eqnarray}
In view of this expression, the matrices $\mu_i$ and $\nu_j$ can be chosen as
\begin{eqnarray}
\mu_1 ~=~0,~~~~~~\mu_2 ~=~ \left(\begin{array}{cc}
\frac{\lambda}{2}  & 0   \\
0 & \frac{\lambda}{2}
\end{array}
\right),~~~~~~~~~~\nu_1 ~=~0,~~~~~~\nu_2 ~=~ \left(\begin{array}{cc}
-\frac{\lambda}{2}  & 0   \\
0 & -\frac{\lambda}{2}
\end{array}
\right).\label{Quantiz.13}
\end{eqnarray}
Clearly, this choice satisfies relation \eqref{Quantiz.5}. Now we can get the sueprcoproducts
\begin{eqnarray}
\Delta \left(\begin{array}{cc}
Q_+    \\
Q_-
\end{array}
\right)&=& \exp \Big\{\left(\begin{array}{cc}
\frac{\lambda}{2} H & 0   \\
0 & \frac{\lambda}{2} H
\end{array}
\right)\Big \} \dot{\otimes} \left(\begin{array}{cc}
Q_+    \\
Q_-
\end{array}
\right)\nonumber\\
&&~~~~~~+ \sigma \Big(\exp\Big\{\left(\begin{array}{cc}
-\frac{\lambda}{2} H & 0   \\
0 & -\frac{\lambda}{2} H
\end{array}
\right)\Big \} \dot{\otimes} \left(\begin{array}{cc}
Q_+    \\
Q_-
\end{array}
\right)\Big). \label{Quantiz.14}
\end{eqnarray}
Suppose that in the classical (anti)commutation relations \eqref{Quantiz.8}, only the composition $\{Q_- , Q_-\}$ is deformed
\begin{equation}\nonumber
\{Q_- , Q_-\}\;=\;\phi(\lambda, H).
\end{equation}
Imposing the conditions $\Delta(\{Q_- , Q_-\}) = \{\Delta Q_- , \Delta Q_-\}$ and
$\Delta \phi(\lambda, H) = \phi(\lambda, 1 \otimes H+ H \otimes 1)$, one easily obtains the relation
\begin{equation}\nonumber
e^{{\lambda} H} \otimes \phi(\lambda, H)+  \phi(\lambda, H)  \otimes e^{-\lambda H} ~=~\phi(\lambda, 1 \otimes H+ H \otimes 1).
\end{equation}
The solution is
\begin{equation}\nonumber
 \phi(\lambda, H)~=~  \frac{\sinh(\lambda {H})}{\lambda}.
\end{equation}
Finally, the results of quantum deformation for the Lie superbialgebra
$\big(({\cal C}^3+{\cal A}) , {\cal C}_{p=1}^{2,\epsilon} \oplus {\cal A}_{1,1}\big)$
are given by the following statement.\\
{\bf Proposition 4:} {\it The  quantum superalgebra which quantizes the
 Lie superbialgebra $\big(({\cal C}^3+{\cal A}) , {\cal C}_{p=1}^{2,\epsilon} \oplus {\cal A}_{1,1}\big)$ has Hopf structure denoted by
${{U}_{_\lambda}}^{\hspace{-1mm}({\cal C}_{p=1}^{2,\epsilon} \oplus {\cal A}_{1,1})}\big(({\cal C}^3+{\cal A})\big)$
and is, respectively,  characterized by
the following  supercoproduct, supercounit and  superantipode \cite{footnote7}
\begin{eqnarray}
\Delta(\hat{Z}) &=& 1 \otimes \hat{Z} + \hat{Z} \otimes 1,~~~~~~~~~~~~~~~~~~~~ ~\Delta(\hat{H}) ~=~ 1 \otimes \hat{H} + \hat{H} \otimes 1, \nonumber\\
\Delta(\hat{Q}_+) &=& e^{\frac{\lambda}{2} {\hat{H}}} \otimes \hat{Q}_+ + \hat{Q}_+ \otimes e^{-\frac{\lambda}{2} \hat{H}},~~~~~~~
\Delta(\hat{Q}_-) ~=~ e^{\frac{\lambda}{2} \hat{H}} \otimes \hat{Q}_- + \hat{Q}_- \otimes e^{-\frac{\lambda}{2} \hat{H}},\label{Quantiz.15}\\
\epsilon(1)&=& 1,~~~~~~~~~~~ \epsilon({\hat{X}}) ~=~0,~~~~~~{
\hat{X}}\in \{\hat{Z}, \hat{H},  \hat{Q}_+, \hat{Q}_-\},~~~~~~~~~~~~~~~~~\label{Quantiz.16}\\
\gamma(\hat{Z}) &=& - \hat{Z},~~~~~~~~~~~~~~~~~~~~~\gamma(\hat{H}) ~=~ -\hat{H}\nonumber\\
\gamma(\hat{Q}_+) &=& - \hat{Q}_+,~~~~~~~~~~~~~~~~~~\gamma(\hat{Q}_-) \;=\;- \hat{Q}_-,\label{Quantiz.17}
\end{eqnarray}
together with the (anti)commutation relations
\begin{equation}
[\hat{Z} , \hat{Q}_-]\;=\;\hat{Q}_+,~~~~~\{\hat{Q}_- , \hat{Q}_-\}\;=\;\frac{\sinh(\lambda \hat{H})}{\lambda},~~~~~
[\hat{H} ~, ~.]\;=\;0,~~~~~[\hat{Q}_+~ , ~.]\;=\;0.
 \label{Quantiz.18}
\end{equation}}
The quantum Casimir belonging to the centre of
${{U}_{_\lambda}}^{\hspace{-1mm}({\cal C}_{p=1}^{2,\epsilon} \oplus {\cal A}_{1,1})}\big(({\cal C}^3+{\cal A})\big)$ (whose classical limit
is \eqref{Quantiz.10})
is generated by
\begin{equation}
{{\cal \hat{C}}_{_{\lambda}}}^{^{(2)}} ~=~ 2 \big(\hat{Z} \frac{\sinh(\lambda \hat{H}\big)}{\lambda} - \hat{Q}_+  \hat{Q}_-).
 \label{Quantiz.19}
\end{equation}
This quantization procedure can be applied to the remaining types of Lie superbialgebras in the
same way. We also quantize the
Lie superbialgebras $\big(({\cal C}^3+{\cal A}) , {\cal C}^{4,\epsilon} \oplus {\cal A}_{1,1}\big)$ and
$\big(({\cal C}^3+{\cal A}) , {\cal C}_{_{p=-1}}^{1,\epsilon} \oplus {\cal A}\big)$. The results are given by the following statements.
\\
{\bf Proposition 5:} {\it The supercoproduct $\Delta$, supercounit $\epsilon$, superantipode $\gamma$
\begin{eqnarray}
\Delta(\hat{Z}) &=& 1 \otimes \hat{Z} + \hat{Z} \otimes 1,~~~~~~~~\Delta(\hat{Q}_+) ~=~ 1 \otimes \hat{Q}_+ + \hat{Q}_+ \otimes e^{-{\lambda} \hat{H}}, \nonumber\\
\Delta(\hat{H}) &=& 1\otimes \hat{H} + \hat{H} \otimes 1,~~~~~~~
\Delta(\hat{Q}_-) ~=~ 1 \otimes \hat{Q}_- + \hat{Q}_- \otimes e^{-{\lambda} \hat{H}} - \hat{Q}_+ \otimes \hat{H}  e^{-{\lambda} \hat{H}},\label{Quantiz.20}\\
\epsilon(1)&=& 1,~~~~~~~~~~~ \epsilon({\hat{X}}) ~=~0,~~~~~~{
\hat{X}}\in \{\hat{Z}, \hat{H},  \hat{Q}_+, \hat{Q}_-\},~~~~~~~~~~~~~~~~~\label{Quantiz.21}\\
\gamma(\hat{Z}) &=& - \hat{Z},~~~~~~~~~~~~~~~~~~~~~~~~~~~~~~~\gamma(\hat{H}) ~=~ -\hat{H} \nonumber\\
\gamma(\hat{Q}_+) &=& - \hat{Q}_{_+} e^{{\lambda} \hat{H}} -\hat{H} \hat{Q}_{_-} e^{{\lambda}\hat{H}},~~~~~~\;~~\gamma(\hat{Q}_-) \;=\;- \hat{Q}_{_-} e^{{\lambda} \hat{H}},\label{Quantiz.22}
\end{eqnarray}
and the (anti)commutation relations
\begin{equation}
[\hat{Z} , \hat{Q}_-]\;=\;\hat{Q}_+,~~~~~~\{\hat{Q}_- , \hat{Q}_-\}\;=\;\frac{1-e^{-2 \lambda \hat{H}}}{2 \lambda},~~~~~
[\hat{H} ~, ~.]\;=\;0,~~~~~[\hat{Q}_+~ , ~.]\;=\;0.
 \label{Quantiz.23}
\end{equation}
determine a Hopf superalgebra denoted by ${{U}_{_{\lambda}}}^{\hspace{-1mm}({\cal C}^{4,\epsilon} \oplus {\cal A}_{1,1})}\big(({\cal C}^3+{\cal A})\big)$
which quantizes the
Lie superbialgebra $\big(({\cal C}^3+{\cal A}) , {\cal C}^{4,\epsilon} \oplus {\cal A}_{1,1}\big)$.}

The deformed Casimir
that commutes with all the generators of the quantum superalgebra , in this case, reads
\begin{equation}
{{\cal \hat{C}}_{_{\lambda}}}^{^{(2)}} ~=~ \frac{1}{\lambda} \hat{Z} (1-e^{-2 \lambda \hat{H}})- 2 \hat{Q}_+  \hat{Q}_-.
\label{Quantiz.23.1}
\end{equation}
\\
{\bf Proposition 6:} {\it The Hopf superalgebra denoted by
${{U}_{_\lambda}}^{\hspace{-1mm}({\cal C}_{_{p=-1}}^{1,\epsilon} \oplus {\cal A})}\big(({\cal C}^3+{\cal A})\big)$
which quantizes the  Lie superbialgebra
$\big(({\cal C}^3+{\cal A}) , {\cal C}_{_{p=-1}}^{1,\epsilon} \oplus {\cal A}\big)$ has,respectively, the supercoproduct, the supercounit
and the superantipode
\begin{eqnarray}
\Delta(\hat{Z}) &=& 1 \otimes \hat{Z} + \hat{Z} \otimes e^{{\lambda} \hat{H}},~~~~~~~~
\Delta(\hat{Q}_+) ~=~ 1 \otimes \hat{Q}_+ + \hat{Q}_+ \otimes 1, \nonumber\\
\Delta(\hat{H}) &=& 1\otimes \hat{H} + \hat{H} \otimes 1,~~~~~~~~~~~
\Delta(\hat{Q}_-) ~=~ 1 \otimes \hat{Q}_- + \hat{Q}_- \otimes e^{-{\lambda} \hat{H}},\label{Quantiz.24}\\
\epsilon(1)&=& 1,~~~~~~~~~~~ \epsilon({\hat{X}}) ~=~0,~~~~~~{
\hat{X}}\in \{\hat{Z}, \hat{H},  \hat{Q}_+, \hat{Q}_-\},~~~~~~~~~~~~~~~~~\label{Quantiz.25}\\
\gamma(\hat{Z}) &=& - \hat{Z} e^{-{\lambda} \hat{H}},~~~~~~~~~~~~~~~~~~~~\gamma(\hat{H}) ~=~ -\hat{H} \nonumber\\
\gamma(\hat{Q}_+) &=& - \hat{Q}_{_+} ,~~~~~~~~~~~~~~~~~~~~~~~~\gamma(\hat{Q}_-) \;=\;- \hat{Q}_{_-} e^{{\lambda} \hat{H}},\label{Quantiz.26}
\end{eqnarray}
with the same (anti)commutation rules and the Casimir element derived in \eqref{Quantiz.23} and \eqref{Quantiz.23.1}, respectively.}

\subsection{Integrable deformation of
 Hamiltonian systems}

Let ${\cal A}$ be a the Hopf superalgebra. Then the pair of $({\cal A} , \Delta)$ is called a supercoalgebra. Given any supercoalgebra
$({\cal A} , \Delta)$ with the corresponding Casimir
element, each of its representations gives rise to a family of completely integrable Hamiltonians \cite{Ballesteros}.
In this subsection, we  construct a deformed quantum integrable Hamiltonian system from the representation of the Lie supercoalgebra
$({\cal C}^3+{\cal A})$ given by Proposition 4, along with the corresponding deformed Casimir element. This system is constructed
on a supersymplectic flat supermanifold of the superdimension-$(4|4)$ as the phase superspace.

Consider $\cal M$ be a supermanifold with a non-degenerate
supersymplectic form $\omega=\frac{(-1)^{^{\Upsilon\Lambda}}}{2}\;\omega_{_{\Upsilon\Lambda}} \;d\Phi^{^{\Upsilon}} \\\wedge
d\Phi^{^{\Lambda}}$ where  $\Phi^{^{\Upsilon}}$ are the local coordinates of ${\cal M}$.
The coordinates $\Phi{^{^\Upsilon}}$
include the bosonic  and the fermionic
coordinates, and the label $\Upsilon$  runs over $\mu=0,\cdots,d_B-1$ and  $\alpha=1,\cdots,d_F$ where $d_B$ and $d_F$
indicate the dimension of the bosonic coordinates
and the fermionic coordinates, respectively.

We define the graded Poisson
bracket structure on supermanifold ${\cal M}$ for two arbitrary functions
$F, G \in C^{\infty}(\Phi^{^{\Upsilon}})$ as \cite{footnote8}
\begin{equation}
\{ F , G\}_{_ {P. B.}}\;=\;\  \frac{F{\overleftarrow{\partial}}}{\partial
\Phi^{^{\Upsilon}}}\;\omega^{^{\Upsilon \Lambda}}\; \frac{{\overrightarrow{\partial}}G}
{\partial
\Phi^{^{\Lambda}}}=(-1)^{^{\Lambda(\Upsilon+|F|)}}\;\omega^{^{\Upsilon \Lambda}}\frac{{\overrightarrow{\partial}}F}
{\partial \Phi^{^{\Upsilon}}}\frac{{\overrightarrow{\partial}}G} {\partial
\Phi^{^{\Lambda}}},\label{integrable:1}
\end{equation}
where $\omega^{^{{\Upsilon\Lambda}}}$ is the superinverse of
$\omega_{_{{\Upsilon\Lambda}}}$. Here we assume that ${\cal M}$ is a flat supermanifold ${\mathbf{R}}_c^4 \times {\mathbf{R}}_a^4$ of
the superdimension-$(4|4)$ with the local coordinates $(q^{\mu}, p_{_\mu}; ~\xi^\alpha, \pi_{_\alpha}), ~({\mu}, \alpha=1, 2)$  and the supersymplectic structure
\begin{equation}
\omega~=~ dq^1 \wedge dp_1 + dq^2 \wedge dp_2 - d\xi^1 \wedge d\pi_1 - d\xi^2 \wedge d\pi_2.  \label{integrable:2}
\end{equation}
Hence, the canonical graded Poisson brackets are given by
\begin{eqnarray}\label{integrable:3}
\{q^{\mu} , p_{_\nu}\}_{_ {P. B,}} &=&- \{p_{_\nu} , q^{\mu}\}_{_ {P. B.}}~=~\delta^{\mu}_{~\nu},~~~~  \{\xi^\alpha , \pi_{_ \beta}\}_{_ {P. B.}}
~=~\{\pi_{_ \beta} ,   \xi^\alpha\}_{_ {P. B.}}~=~\delta^\alpha_{~\beta}, \nonumber\\
\{q^{\mu} , \xi^\alpha\}_{_ {P. B.}} &=&\{p_{_\mu} , \pi_{_ \alpha}\}_{_ {P. B.}}~=~\{\xi^\alpha , p_{_\mu}\}_{_ {P. B.}}
~=~\{q^{\mu} ,   \pi_{_ \alpha}\}_{_ {P. B.}}~=~0,
\end{eqnarray}
where $p_{_\mu} = -\frac{\partial}{\partial q^{\mu}}$ and $\pi_{_ \alpha} = \frac{\overrightarrow{\partial}}{\partial\xi^\alpha}$ are the
conjugate to $q^{\mu}$ and $\xi^\alpha$ momentums, respectively.

We now  consider that the Lie superalgebra $\g$ is realized by means of smooth functions on the phase superspace
${\mathbf{R}}_c^4 \times {\mathbf{R}}_a^4$  with the local coordinates $(q^{\mu}, p_{_\mu}; ~\xi^\alpha, \pi_{_\alpha})$
\begin{eqnarray}\label{integrable:4}
S(X_i)~=~X_i(q^{\mu}, p_{_\mu}; ~\xi^\alpha, \pi_{_\alpha}).
\end{eqnarray}
This means  that under the canonical graded Poisson bracket
\begin{eqnarray}\label{integrable:5}
\{ F , G\}_{_ {P. B.}}\;=\;\  \sum_{\mu=1}^{2} \Big(\frac{{\partial}F}{\partial q^{\mu}} \frac{{\partial}G}{\partial
p_{_\mu}}- \frac{{\partial}F}{\partial p_{_\mu}} \frac{{\partial}G}{\partial
q^{\mu}}\Big)-(-1)^{|F|}  \sum_{\alpha=1}^{2} \Big(\frac{\overrightarrow{\partial}F}{\partial
\xi^\alpha} \frac{\overrightarrow{\partial}G}{\partial \pi_{_ \alpha}}+ \frac{\overrightarrow{\partial}F}{\partial
\pi_{_ \alpha}} \frac{\overrightarrow{\partial}G}{\partial \xi^\alpha}\Big),
\end{eqnarray}
the dynamical variables $S(X_i)$ close the initial Lie superalgebra
\begin{equation}
\{S({X_i})\;,\;S({X_j})\}_{_ {P. B.}}\;=\;f_{~ij}^{k}\;S({X_k}).\label{integrable:6}
\end{equation}
Clearly, using relation \eqref{integrable:5}, \eqref{integrable:6} is reduced to the following system of partial differential equations (PDEs)
\begin{eqnarray}\label{integrable:7}
&&\sum_{\mu=1}^{2} \Big(\frac{{\partial}S({X_i})}{\partial q^{\mu}} \frac{{\partial}S({X_j})}{\partial
p_{_\mu}}- \frac{{\partial}S({X_i})}{\partial p_{_\mu}} \frac{{\partial}S({X_j})}{\partial
q^{\mu}}\Big)~~~~~~~~\nonumber\\
&&~~~~~-(-1)^{i}  \sum_{\alpha=1}^{2} \Big(\frac{\overrightarrow{\partial}S({X_i})}{\partial
\xi^\alpha} \frac{\overrightarrow{\partial}S({X_j})}{\partial \pi_{_ \alpha}}+ \frac{\overrightarrow{\partial}S({X_i})}{\partial
\pi_{_ \alpha}} \frac{\overrightarrow{\partial}S({X_j})}{\partial \xi^\alpha}\Big) - f_{~ij}^{k}\;S({X_k})~=~0.
\end{eqnarray}
Here the generators ${X_i}$ are $Z, H, Q_+$ and $Q_-$, and $f_{~ij}^{k}$ stand for the structure constants of $({\cal C}^3+{\cal A})$.
Note that two different  realizations \eqref{integrable:4} (two different solutions of (PDEs)  \eqref{integrable:7}) will be equivalent if there exists a canonical transformation that maps one into the other.
A convenient realization linked to $({\cal C}^3+{\cal A})$ (a solution for the above system of (PDEs) ) given by
\begin{eqnarray}\label{integrable:8}
S({Z})&=& -q^1 p_2 -\xi^1 \pi_2, ~~~~~~~~~~~~S({H}) ~=~ -q^1 p_2 +\xi^1 \pi_2, \nonumber\\
S({Q_+}) &=& q^1 \pi_2, ~~~~~~~~~~~~~~~~~~~~~~S({Q_-}) ~=~ -\xi^1 p_2+ \frac{1}{2}(q^1 \pi_1+ q^2 \pi_2).
\end{eqnarray}
This realization  can be easily deformed:
\begin{eqnarray}\label{integrable:9}
S({\hat{Z}})&=& -q^1 p_2 -\xi^1 \pi_2,~~~~~~~~~~S({\hat{H}}) ~=~ - \frac{1}{\lambda} \sinh(\lambda q^1 p_2) +\xi^1 \pi_2 \cosh(\lambda q^1 p_2), \nonumber\\
S({\hat{Q}_+}) &=& q^1 \pi_2,~~~~~~~~~~~~~~~~~~~~
S({\hat{Q}_-}) ~=~ -\frac{\xi^1}{\lambda q^1}\sinh(\lambda q^1 p_2) + \frac{1}{2}(q^1 \pi_1+ q^2 \pi_2).
\end{eqnarray}
These phase superspace functions close a quantum superalgebra $({\cal C}^3+{\cal A})$ \eqref{Quantiz.18} under the canonical
graded Poisson bracket \eqref{integrable:5}.
Under the realization \eqref{integrable:8}, the undeformed Casimir element of $({\cal C}^3+{\cal A})$ (given by relation \eqref{Quantiz.10})
is represented by
\begin{eqnarray}\label{integrable:10}
S({\cal C}^{^{(2)}})~=~2(q^1)^2 (p_{_2})^2 -2 q^1 p_{_2} \xi^1 \pi_{_2} + (q^1)^2 \pi_1 \pi_{_2}.
\end{eqnarray}
One can easily show that $S({\cal C}^{^{(2)}})$ will always be in involution with the functions $S({X_i})$ \eqref{integrable:8}
in such a way that it cab be
considered as a common constant of motion. Therefore, a particular subset of undeformed integrable Hamiltonian can be found  by setting
\begin{eqnarray}\label{integrable:11}
{\cal H} &=& S({\cal C}^{^{(2)}}) + {\cal F}\big(S({{H}})\big)\nonumber\\
&=& 2(q^1)^2 (p_{_2})^2 -2 q^1 p_{_2} \xi^1 \pi_{_2} + (q^1)^2 \pi_1 \pi_{_2} + {\cal F}\big(-q^1 p_{_2} +\xi^1 \pi_{_2}\big).
\end{eqnarray}
where ${\cal F}\big(S({{H}})\big)$ is any smooth function of the dynamical variable $S({{H}})$.

We recall that the deformed Casimir element for the Hopf superalgebra ${{U}_{_\lambda}}^{\hspace{-1mm}({\cal C}_{p=1}^{2,\epsilon} \oplus {\cal A}_{1,1})}\big(({\cal C}^3+{\cal A})\big)$ has given by relation \eqref{Quantiz.19}. In terms of the deformed realization \eqref{integrable:9},
it reads
\begin{eqnarray}\label{integrable:12}
S({{\cal \hat{C}}_{_{\lambda}}}^{^{(2)}})~=~ 2 \Big\{S(\hat{Z}) \frac{\sinh(\lambda S(\hat{H})\big)}{\lambda} - S(\hat{Q}_+)  S(\hat{Q}_-)\Big\}.
\end{eqnarray}
If we consider the deformed (anti)commutation rules \eqref{Quantiz.18} as graded Poisson brackets, then we get that \eqref{integrable:12}
Poisson-commutes with any function of  \eqref{integrable:9}. An example of the deformed Hamiltonian is provided by
\eqref{integrable:11} where the phase superspace functions are now replaced by their deformed counterparts
\begin{eqnarray}\label{integrable:13}
{\cal \hat{H}}_{_\lambda}~=~ S({{\cal \hat{C}}_{_{\lambda}}}^{^{(2)}})+{\cal F}\big(S({{\hat H}})\big),
\end{eqnarray}
by construction, ${\cal \hat{H}}_{_\lambda}$ is in involution with the deformed functions of \eqref{integrable:9}. Of course,
in order to obtain the Casimir as a Hamiltonian, we should consider ${\cal \hat{H}}_{_\lambda} \equiv  S({{\cal \hat{C}}_{_{\lambda}}}^{^{(2)}})$; then,
${\cal F}\big(S({{\hat H}})\big)$ can be taken as the remaining integral of the motion in involution.


\section{Conclusion}

In summary, as mentioned in the Introduction, the importance of the classification of
Lie superbialgebras $({ \cal C}^3 + { \cal A})$ lies in the fact that it helps us to obtain a hierarchy of $({ C}^3 + { A})$ WZW models related to the
super Poisson-Lie T-duality \cite{ER8}. We have performed a complete classification of
Lie superbialgebras $({ \cal C}^3 + { \cal A})$ and obtained 31 families of inequivalent Lie superbialgebra
structures on  $({ \cal C}^3 + { \cal A})$ in such a way that all results are new and applicable.
We have also classified all corresponding coboundary Lie superbialgebras
(triangular or quasi-triangular)
with  coboundary duals and their corresponding classical $r$-matrices.
The generalization of the Lyakhovsky and Mudrov formalism \cite{Lyakhovsky}  to Lie superalgebras was firstly done in Ref. \cite{geometry}.
Making use of this formalism, we have obtained new quantum deformations of $({ \cal C}^3 + { \cal A})$. In this way, we have gotten
the Hopf superalgebras which quantize the
Lie superbialgebras $\big(({\cal C}^3+{\cal A}) , {\cal C}_{p=1}^{2,\epsilon} \oplus {\cal A}_{1,1}\big)$, $\big(({\cal C}^3+{\cal A}) , {\cal C}^{4,\epsilon} \oplus {\cal A}_{1,1}\big)$ and $\big(({\cal C}^3+{\cal A}) , {\cal C}_{_{p=-1}}^{1,\epsilon} \oplus {\cal A}\big)$ of Table 3.
Finally, as an application of these quantum deformations, a deformed integrable Hamiltonian system  has been constructed by the representation of the Hopf superalgebra ${{U}_{_\lambda}}^{\hspace{-1mm}({\cal C}_{p=1}^{2,\epsilon} \oplus {\cal A}_{1,1})}\big(({\cal C}^3+{\cal A})\big)$.


\bigskip
{\it Acknowledgments}:~ This work has been supported by Iran National Science Foundation:INSF under research fund No. 94/s/41258.



\begin{thebibliography}{}
\bibitem{footnote3}
{We call $(-1)^{|x|}$ the parity of $x$. Here and in the following we identify grading of indices by the same indices in the power of $(-1)$, i.e., we put $(-1)^x$ instead of  $(-1)^{|x|}$; this notation
has used by Dewitt in \cite{D}.}
\bibitem{footnote7}
Note that here we identify deformed generators by putting hat sign over them.
\end{thebibliography}
\end{document}